\documentclass[12pt,a4paper]{article}
\usepackage{amssymb}
\usepackage{graphicx}
\usepackage{graphics}
\usepackage{epsfig}
\usepackage{amsmath}
\usepackage{url}
\usepackage{multicol}
\usepackage{float}
\usepackage{empheq}
\begin{document}

\begin{center}
{\Large \bf Theory of $U(1)$ gauged Q-balls revisited}

\vspace{4mm}

I.\,E.\;Gulamov$^{a,b}$, E.\,Ya.\;Nugaev$^c$,
M.\,N.\;Smolyakov$^b$\\
\vspace{0.5cm} $^a${\small{\em Physics Department, Lomonosov
Moscow State University,
}}\\
{\small{\em 119991, Moscow, Russia}}\\
$^b${\small{\em Skobeltsyn Institute of Nuclear Physics, Lomonosov
Moscow State University,
}}\\
{\small{\em 119991, Moscow, Russia}}\\
$^c${\small{\em Institute for Nuclear Research of the Russian
Academy
of Sciences,}}\\
{\small{\em 60th October Anniversary prospect 7a, 117312, Moscow,
Russia }}
\end{center}

\begin{abstract}
In this paper, the main properties of (3+1)-dimensional $U(1)$
gauged Q-balls are examined. In particular, it is shown that the
relation $\frac{dE}{dQ}=\omega$ holds for such gauged Q-balls in
the general case. As a consequence, it is shown that the
well-known estimate for the maximal charge of stable gauged
Q-balls was derived by means of an inconsistent procedure and can
not be considered as correct. A simple method for obtaining the
main characteristics of gauged Q-balls using only the nongauged
background solution for the scalar field in the case, when the
back-reaction of the gauge field on the scalar field is small and
the linearized theory can be used, is proposed. The criteria of
applicability of the linearized theory, which do not reduce to the
demand of the smallness of the coupling constant, are established.
Some interesting properties of gauged Q-balls, as well as the
advantages of the proposed method, are demonstrated by the example
of two models, admitting, in the linear approximation in the
perturbations, exact analytic solutions for gauged Q-balls.
\end{abstract}

\section{Introduction}
Non-topological solitons in a theory of complex scalar field with
global $U(1)$ symmetry, proposed in \cite{Rosen0} and known as
Q-balls \cite{Coleman:1985ki}, are widely discussed in the
literature. A simplest generalization of Q-balls to the gauged
case, i.e., from the global $U(1)$ symmetry to the gauge $U(1)$
symmetry, is straightforward. Although the existence of gauged
Q-balls was put in question in the well-known paper
\cite{Coleman:1985ki}, there are some papers devoted to this
subject. The most known paper is \cite{Lee:1988ag}, where gauged
Q-balls were examined analytically and numerically (for
simplicity, from here on, we call $U(1)$ gauged Q-balls ``gauged
Q-balls'', unless otherwise stated). To our knowledge, for the
first time, an analysis of what is now called gauged Q-balls was
made in \cite{Rosen}. In this remarkable paper not only were the
conditions for the existence of such Q-balls derived, but also the
case of small coupling of the gauge field to the scalar field was
discussed, the corresponding linearized equations of motion were
obtained, and even an approximate solution to these equations was
found. One can also recall paper \cite{Lee:1991bn}, where gauged
non-topological soliton solutions were obtained numerically. The
scalar field part of the model of \cite{Lee:1991bn} coincides with
the two-fields model proposed in \cite{Friedberg:1976me}; that is
why the gauged non-topological soliton solution found in
\cite{Lee:1991bn} is not a gauged Q-ball in the sense of Coleman's
definition of Q-balls \cite{Coleman:1985ki}, although it is of the
same kind. Another interesting paper is \cite{Arodz:2008nm}, in
which not only numerical but also approximate analytic solutions
for gauged Q-balls and Q-shells were obtained. For a certain class
of scalar field potentials and for a sufficiently weak interaction
between the scalar field and the $U(1)$ gauge field, the existence
of gauged Q-balls was proven in \cite{BF,BF1} in a mathematically
rigorous way. A solution for a gauged Q-ball in such a case of a
weak coupling of the $U(1)$ gauge field to the scalar field (i.e.,
an exact solution in the linear approximation above the background
solution) was recently found in \cite{Dzhunushaliev:2012zb} in the
gauged version of the model proposed in \cite{Rosen:1969ay}; the
solution for the case in which the coupling in this model is not
weak was studied in \cite{Dzhunushaliev:2012zb} numerically.

Meanwhile, we think that there is a lack of understanding of the
physical properties of $U(1)$ gauged Q-balls, so in this paper we
present some results concerning both the general properties of
gauged Q-balls and the particular case of the small back reaction
of the gauge field. The paper is organized as follows. In
Section~2 we present the general setup and introduce the notations
that will be used throughout the paper. In Section~3 we study the
main properties of gauged Q-balls; in particular, we prove that
the relation $\frac{dE}{dQ}=\omega$ holds for {\em any} gauged
Q-ball. We also discuss different issues concerning the stability
of such Q-balls and show that the well-known statement about the
existence of a maximal charge of stable gauged Q-balls, which was
made in \cite{Lee:1988ag}, is incorrect. In Section~4 we
thoroughly examine the case in which the back-reaction of the
gauge field on the scalar field is small. We propose a very useful
method for studying the gauged Q-ball properties {\em without}
solving the whole system of linearized equations of motion for the
fields. The resulting compact formulas allow us to simplify the
calculations considerably. We show that the small parameter of the
theory does not coincide with $e^2$ in the general case (here, $e$
is the coupling constant of the gauge field to the scalar field),
which implies that the fulfillment of the condition $e^{2}\ll 1$
does not guarantee that the linearized theory can be used for
calculations. These results are illustrated by examples of two
models providing, in the linear approximation in the
perturbations, exact analytic solutions for gauged Q-balls.

\section{Setup}
We consider the action, describing the simplest $U(1)$ gauge
invariant four-dimensional scalar field theory, in the form
\begin{equation}\label{action}
S=\int
d^4x\left((\partial^{\mu}\phi^{*}-ieA^{\mu}\phi^{*})(\partial_{\mu}\phi+ieA_{\mu}\phi)-V(\phi^{*}\phi)-\frac{1}{4}F_{\mu\nu}F^{\mu\nu}\right)
\end{equation}
and take the standard spherically symmetric ansatz for the fields
describing a gauged Q-ball:
\begin{eqnarray}\label{ans1}
\phi(t,\vec x)&=&\textrm{e}^{i\omega t}f(r),\qquad
f(r)|_{r\to\infty}\to 0, \qquad \frac{df(r)}{dr}\biggl|_{r=0}=0,\\
\label{ans2}A_{0}(t,\vec x)&=&A_{0}(r),\qquad\,\,
A_{0}(r)|_{r\to\infty}\to 0,\qquad
\frac{dA_{0}(r)}{dr}\biggl|_{r=0}=0,\\ \label{ans3} A_{i}(t,\vec
x)&\equiv &0,
\end{eqnarray}
where $r=\sqrt{\vec x^{2}}$ and $f(r)$, $A_{0}(r)$ are real
functions. We suppose that the function $f(r)$ has no nodes and
$f(0)>0$.

Taking into account (\ref{ans1})--(\ref{ans3}), we can use the
effective action
\begin{equation}\label{effaction}
S_{\textrm{eff}}=\int
d^3x\left((\omega+g)^{2}f^2-\partial_{i}f\partial_{i}f-V(f)+\frac{1}{2e^2}\partial_{i}g\partial_{i}g\right),
\end{equation}
where $g=eA_{0}$, $V(f)=V(\phi^{*}\phi)$ (Eq.~(\ref{ans1}) implies
that $\phi^{*}\phi=f^{2}$), instead of
(\ref{action}).\footnote{Though the Q-ball solution is supposed to
be spherically symmetric, sometimes it is convenient to keep the
coordinates $x^{i}$ and the corresponding volume element $d^{3}x$,
especially in the calculations for which the spherical symmetry of
the fields is not required.} For the scalar field potential, the
conditions
\begin{equation}
V(0)=0,\qquad \frac{dV}{df}\biggl|_{f=0}=0
\end{equation}
are supposed to fulfill. It should be noted that gauged Q-balls in
theories with $V(f)\equiv 0$ or $V(f)=M^2f^2$ do not exist; see,
for example, \cite{Rosen}.

The equations of motion, following from effective action
(\ref{effaction}), take the form
\begin{eqnarray}\label{eqg1}
2e^2(\omega+g)f^2=\Delta g,\\
\label{eqg2} 2(\omega+g)^2f+2\Delta f-\frac{dV}{df}=0,
\end{eqnarray}
where $\Delta=\sum\limits_{i=1}^{3}\partial_{i}\partial_{i}$. We
define the charge of a gauged Q-ball as
\begin{equation}\label{chargedef}
Q=2\int d^3x(\omega+g)f^2.
\end{equation}
Note, that the physical charge is
\begin{equation}
Q_{phys}=eQ,
\end{equation}
but for convenience, below we will use the charge $Q$ defined by
(\ref{chargedef}), not $Q_{phys}$.

It was shown in \cite{Lee:1988ag} that for a gauged Q-ball
solution the sign of $\omega+g$ always coincides with the sign of
$\omega$. Because of the symmetry $\omega\to-\omega$, $g\to-g$ of
the equations of motion, without loss of generality for simplicity
we can consider $\omega\ge 0$. In this case, according to
(\ref{chargedef}), for $\omega>0$ we get $Q>0$. As it was shown in
\cite{Rosen}, $g\equiv 0$ for $\omega=0$ and only a purely static
solution for the scalar field can exist, so our choice $\omega\ge
0$ implies $Q\ge 0$.

The energy of a gauged Q-ball at rest is defined by
\begin{equation}\label{energydef}
E=\int
d^3x\left((\omega+g)^2f^2+\partial_{i}f\partial_{i}f+V(f)+\frac{1}{2e^2}\partial_{i}g\partial_{i}g\right).
\end{equation}

\section{Some general properties of $U(1)$ gauged Q-balls}
\subsection{$\frac{dE}{dQ}$ for gauged Q-balls} It is well known
that for ordinary (nongauged) Q-balls the relation
$\frac{dE}{dQ}=\omega$ holds. We have failed to find any note
about the validity of this or an analogous relation for Abelian
gauged Q-balls in the literature on this subject. So, below we
present a simple proof of the fact that for gauged Q-balls of form
(\ref{ans1})--(\ref{ans3}) in a theory described by action
(\ref{action}) the relation $\frac{dE}{dQ}=\omega$ also holds.

It is reasonable to suppose that the only parameter, which
characterizes the charge and the energy for given $V(f)$ and $e$,
is $\omega$. Thus, differentiating the energy (\ref{energydef})
with respect to $\omega$, we get
\begin{eqnarray}\nonumber
\frac{dE}{d\omega}&=&\int\left(2\frac{d(\omega+g)}{d\omega}(\omega+g)f^2+2(\omega+g)^2f\frac{df}{d\omega}+
2\partial_{i}f\partial_{i}\frac{df}{d\omega}+\frac{dV}{df}\frac{df}{d\omega}\right.\\
\nonumber&+&\left.\frac{1}{e^2}\partial_{i}g\partial_{i}\frac{dg}{d\omega}\right)d^3x=
\int\left(2\frac{d(\omega+g)}{d\omega}(\omega+g)f^2+2(\omega+g)^2f\frac{df}{d\omega}\right.\\
\nonumber &+&\left.\left(-
2\Delta f+\frac{dV}{df}\right)\frac{df}{d\omega}+\frac{1}{e^2}\partial_{i}g\partial_{i}\frac{dg}{d\omega}\right)d^3x\\
\label{dedw} &=&
\int\left((\omega+g)\left(2\frac{d(\omega+g)}{d\omega}f^2+4(\omega+g)f\frac{df}{d\omega}\right)+\frac{1}{e^2}\partial_{i}g\partial_{i}\frac{dg}{d\omega}\right)d^3x,
\end{eqnarray}
where we have used Eq.~(\ref{eqg2}). For convenience, let us use
the notation $q=2(\omega+g)f^{2}$. Equation (\ref{dedw}) can be
rewritten as
\begin{eqnarray}\label{dedw1A}
\frac{dE}{d\omega}=
\int\left((\omega+g)\frac{dq}{d\omega}+\frac{1}{e^2}\partial_{i}g\partial_{i}\frac{dg}{d\omega}\right)d^3x=\omega\frac{dQ}{d\omega}+
\int\left(g\frac{dq}{d\omega}+\frac{1}{e^2}\partial_{i}g\partial_{i}\frac{dg}{d\omega}\right)d^3x,
\end{eqnarray}
where, according to (\ref{chargedef}), $Q=\int qd^3x$.
Eq.~(\ref{eqg1}) implies that
$\frac{dq}{d\omega}=\frac{1}{e^2}\Delta\frac{dg}{d\omega}$.
Substituting it into Eq.~(\ref{dedw1A}), we arrive at
\begin{eqnarray}\label{dedw2A}
\frac{dE}{d\omega}=\omega\frac{dQ}{d\omega}+
\frac{1}{e^2}\int\left(g\Delta
\frac{dg}{d\omega}+\partial_{i}g\partial_{i}\frac{dg}{d\omega}\right)d^3x.
\end{eqnarray}
The integral in (\ref{dedw2A}) is equal to zero, which can be
easily seen by performing integration by parts (since
$g|_{r\to\infty}\sim\frac{Q}{r}$ is assumed for gauged Q-balls and
consequently
$\frac{dg}{d\omega}\bigl|_{r\to\infty}\sim\frac{dQ/d\omega}{r}$,
the surface term, arising when an integration by parts is
performed, obviously vanishes). Thus, we get
$\frac{dE}{d\omega}=\omega\frac{dQ}{d\omega}$, which leads to
\begin{eqnarray}\label{dEdQgauged}
\frac{dE}{dQ}=\omega
\end{eqnarray}
for $\frac{dQ}{d\omega}\ne 0$. We stress that the fulfillment of
(\ref{dEdQgauged}) is the general property inherent to any $U(1)$
gauged Q-ball. As for the points at which $\frac{dQ}{d\omega}=0$
(and, consequently, $\frac{dE}{d\omega}=0$), they correspond to
the cusps on the $E(Q)$ diagram (like the cusps in
Fig.~\ref{EQ_Rosen} and Fig.~\ref{EQ} of Section~4), indicating
the existence of a (locally) minimal or (locally) maximal charge.
The cusps also separate different branches of the $E(Q)$
dependence.

\subsection{Stability of gauged Q-balls}
In the absence of interactions with fermions there are three types
of stability of Q-balls. They are: the stability with respect to
decay into free particles (i.e., quantum mechanical stability);
the stability against decay into Q-balls with smaller charges
(i.e., against fission) and the classical stability (the stability
with respect to small perturbations of fields). Here, we will not
consider the classical stability, because in the general case its
consistent study for gauged Q-balls is a rather complicated task
and lies beyond the scope of this paper. Here we will consider
only the quantum mechanical stability and stability against
fission.

\subsubsection*{Quantum mechanical stability and maximal charge of gauged Q-balls}
We start with the stability with respect to decay into free scalar
particles. Suppose that there exist free particles of mass
$M=\sqrt{\frac{1}{2}\frac{d^{2}V}{df^{2}}\bigl|_{f\equiv 0}}$ in
the theory at hand (all the reasonings presented below are based
on the assumption that there are no extra fields except those in
action (\ref{action}); otherwise, the situation can be more
complicated). In this case, the criterion for Q-ball stability
looks very simple,
\begin{equation}\label{stab1}
E(Q)<MQ,
\end{equation}
where $E(Q)$ is the energy of a Q-ball with the charge $Q$.

It is necessary to note that, as it was shown in
\cite{Lee:1988ag,Rosen}, the inequality $$\omega<M$$ should hold
for a Q-ball in such a theory. Indeed, the existence of free
scalar particles of mass $M$ implies that the relevant scalar
field part of the action has the form
\begin{equation}\label{quadact}
S_{\textrm{scalar}}\approx\int
d^4x\left(\partial^{\mu}\phi^{*}\partial_{\mu}\phi-M^{2}\phi^{*}\phi\right)
\end{equation}
for small values of $\phi^{*}\phi$. The latter means that for any
Q-ball in this theory, satisfying (\ref{ans1}) and (\ref{ans2}),
inequality $\omega<M$ must hold; otherwise, the corresponding
solution to equation (\ref{eqg2}) does not fall off at infinity
rapidly enough to ensure the finiteness of the Q-ball charge and
energy.

An interesting observation is that gauged Q-balls (as well as
nongauged Q-balls) can not {\em emit} free scalar particles, they
can only {\em decay} into such particles. Indeed, since the charge
of a free particle in the theory at hand is $Q_{p}=1$, for a
Q-ball, we get
\begin{equation}
E(Q+N)=E(Q)+\int\limits_{Q}^{Q+N}\frac{dE}{d\tilde Q}d\tilde
Q<E(Q)+M\int\limits_{Q}^{Q+N}d\tilde Q=E(Q)+MN,
\end{equation}
where $N$ stands for the number of emitted particles. We see that
the emission of $N>0$ free scalar particles is energetically
forbidden.\footnote{This reasoning works only for Q-balls from
{\em the same} branch of the $E(Q)$ dependence, transitions
between Q-balls from different branches (like those in
Fig.~\ref{EQ} of Section~4) with the emission of free scalar
particles, or/and antiparticles and vector particles (photons) are
not energetically forbidden in the general case.} One can easily
show that an analogous emission of scalar antiparticles (i.e.,
particles with $Q_{p}=-1$) is also energetically forbidden.

It this connection we would like to comment on the method of
derivation of the maximal charge of stable gauged Q-balls,
presented in \cite{Lee:1988ag}. Although the corresponding
estimates for the maximal charge were obtained within the
particular model of \cite{Lee:1988ag}, they are used in many
papers concerning gauged Q-balls. It is stated in
\cite{Lee:1988ag} that for a charge $Q$, such that for a Q-ball of
this charge the inequality $\frac{dE}{dQ}>M$ (in our notations)
holds, it is energetically favorable to have a Q-ball with the
charge $Q_{max}$ and $Q-Q_{max}$ free scalar particles. The
maximal charge $Q_{max}$ is defined as a solution to equation
$\frac{dE}{dQ}=M$. As it was shown above, for any gauged Q-ball in
a theory with
$\frac{dV(\phi^{*}\phi)}{d(\phi^{*}\phi)}\bigl|_{\phi^{*}\phi=0}=M^2>0$
the inequality $\frac{dE}{dQ}=\omega<M$ holds, and Q-balls with
$\frac{dE}{dQ}\ge M$ can never exist (contrary to what was stated
in \cite{Lee:1988ag}). If any approximate solution leads to the
existence of gauged Q-balls with $\frac{dE}{dQ}\ge M$ in a theory
admitting the existence of free particles of mass $M$, such an
approximate solution is not valid. Thus, the procedure used in
\cite{Lee:1988ag} for estimating the value of the maximal charge
of stable gauged Q-balls contradicts the main properties of gauged
Q-balls and can not be considered as correct, as well as the
consequent statement about the existence of the maximal
charge.\footnote{In \cite{Loginov:2012zz} it was observed that
Q-balls can not emit free particles of mass $M$ if the condition
$\frac{dE}{dQ}=\omega<M$ is fulfilled.}

Of course, stable gauged Q-balls with maximal charges may exist.
As in the nongauged case (see, for example,
\cite{Gulamov:2013ema,MarcVent}), the existence of the maximal
charge in the gauged case can be determined by the form of the
scalar field potential or by the values of the model parameters,
such a (locally) maximal charge corresponds to a cusp in the
$E(Q)$ dependence. Explicit examples, which will be presented in
Section~4, clearly demonstrate it. Meanwhile, there are many
models with the charge of an absolutely stable nongauged Q-ball
(classically stable, quantum mechanically stable and stable
against fission) not bound from above; see, for example,
\cite{Gulamov:2013ema,Theodorakis:2000bz}. So, we do not see any
evident physical reason why it should not be so for gauged
Q-balls. In this connection, we would like to comment on the
common belief that the Coulomb repulsion makes a gauged Q-ball
with some large charge unstable. Indeed, the repulsion due to the
gauge field exists. But let us look at Eq.~(\ref{eqg1}), which can
be rewritten as
\begin{eqnarray}
\Delta g-2e^2f^2g=2e^2\omega f^2.
\end{eqnarray}
This equation implies that the gauge field inside a Q-ball is
effectively massive, which dilutes the strength of the repulsion
considerably in comparison with the Coulomb long-range repulsion.
Thus, although its influence on the stability of gauged Q-balls
can not be ignored, it should be reconsidered more accurately.

\subsubsection*{Stability against fission}
Now, we turn to examining the stability against fission. For
ordinary (nongauged) Q-balls the corresponding stability criterion
takes the form
\begin{equation}\label{q-decay-crit}
d^{2}E/dQ^{2}<0.
\end{equation}
If $E(0)$=0, then (\ref{q-decay-crit}) clearly leads to
\begin{equation}\label{q-decay2}
E(Q_{1})+E(Q_{2})> E(Q_{1}+Q_{2}),
\end{equation}
which implies that Q-ball fission is energetically forbidden. But
in many models the $d^{2}E/dQ^{2}<0$ branches of the $E(Q)$
dependence are such that there exists a minimal charge $Q_{min}\ne
0$: $E(Q_{min})=E_{min}\ne 0$. In this case, one can try to
redefine the function $E(Q)$ in the region $[0,Q_{min}]$ in order
to get a continuous and differentiable auxiliary function
$E_{aux}(Q)$: $E_{aux}(0)=0$, $E_{aux}(Q)$ is a monotonically
increasing function for $Q>0$, $d^{2}E_{aux}(Q)/dQ^{2}<0$, and
$E_{aux}(Q)=E(Q)$ for $Q\ge Q_{min}$. If it is possible to
construct such a function $E_{aux}(Q)$, then inequality
(\ref{q-decay2}) is valid for $Q_{1}, Q_{2}\ge 0$ and,
consequently, for $Q_{1}, Q_{2}\ge Q_{min}$. Of course, such
reasonings apply for gauged Q-balls too.

It was shown in \cite{Gulamov:2013ema} that the necessary
condition for the existence of the function $E_{aux}(Q)$ is
$\frac{E(Q_{min})}{Q_{min}}>\omega_{min}=\frac{dE}{dQ}\bigl|_{Q=Q_{min}}$.
For nongauged Q-balls this relation always holds because the
equality
\begin{equation}\label{EomegaQbigger}
E=\omega Q+\frac{2}{3}\int d^{3}x\partial_{i}f\partial_{i}f
\end{equation}
holds for nongauged Q-balls, leading to
$E(Q_{min})=\omega_{min}Q_{min}+\frac{2}{3}\int
d^{3}x\partial_{i}f\partial_{i}f>\omega_{min}Q_{min}$.

For $U(1)$ gauged Q-balls one obtains \cite{Lee:1988ag}
\begin{equation}\label{EomegaQbigger1}
E=\omega Q+\int
d^{3}x\left(\frac{2}{3}\partial_{i}f\partial_{i}f-\frac{1}{3e^2}\partial_{i}g\partial_{i}g\right).
\end{equation}
This relation can be easily derived by applying the scale
transformation technique of \cite{Derrick} to effective action
(\ref{effaction}), substituting the result into (\ref{energydef})
(to exclude $V(f)$) and using equation of motion (\ref{eqg1}). One
sees that, contrary to the case of ordinary Q-balls
(\ref{EomegaQbigger}), the integral in (\ref{EomegaQbigger1}) is
not positive-definite. Thus, at this stage, we can not make any
conclusion about the stability against fission for gauged Q-balls
with $d^{2}E/dQ^{2}<0$ in the general case. Of course, one may
simply check that $\frac{E(Q_{min})}{Q_{min}}>\omega_{min}$ for a
particular gauged Q-ball solution. For example, if there exist
free scalar particles of mass $M$ in the theory under
consideration and the part of the $E(Q)$ dependence with
$d^{2}E/dQ^{2}<0$, which we are interested in, starts at
$Q=Q_{min}\ne 0$ and $E(Q_{min})\ge MQ_{min}$ (as we will see
below, it holds at least for one example that will be discussed in
our paper later; see also \cite{Lee:1991bn} where this situation
is realized), then Q-balls from this branch are stable against
fission. Indeed, $\omega_{min}<M\le\frac{E(Q_{min})}{Q_{min}}$,
the latter means that, according to \cite{Gulamov:2013ema}, we
always can construct an auxiliary function $E_{aux}(Q)$ possessing
the properties presented above. An explicit example of the
function $E_{aux}(Q)$ can be found in Appendix~A.

Here, we discuss another possibility. Suppose we have a gauged
Q-ball solutions in a model with $e=e_{x}$ (without loss of
generality, we consider $e_{x}> 0$) for the region of frequencies
$\omega$ we are interested in. Let us also suppose that there
exist nongauged Q-ball solutions in this model with $e=0$ for the
same region of frequencies. In this case, one may assume that
gauged Q-ball solutions exist for any $0<e<e_{x}$. This assumption
is not obvious, and we can not justify it in a mathematically
rigorous way for the general case, meanwhile, as we will see
explicitly in the next section, it is valid at least when the
back-reaction of the gauge field is small (this fact was also
proven in \cite{BF} in a mathematically rigorous way for a certain
class of the scalar field potentials). If the conditions presented
above are fulfilled, then for a gauged Q-ball in a theory with
$e=e_{x}$, the inequality
$$
E>\omega Q
$$
holds. To show it, let us take the energy (\ref{energydef}) and
differentiate it with respect to the coupling constant $e$ while
keeping $\omega$ fixed. Performing calculations analogous to those
made in (\ref{dedw}) and (\ref{dedw1A}) we arrive at
\begin{eqnarray}\label{dede1}
\frac{dE}{de}=\omega\frac{dQ}{de}+
\int\left(g\frac{dq}{de}-\frac{1}{e^3}\partial_{i}g\partial_{i}g+\frac{1}{e^2}\partial_{i}g\partial_{i}\frac{dg}{de}\right)d^3x.
\end{eqnarray}
Eq.~(\ref{eqg1}) implies that
$\frac{dq}{de}=\frac{1}{e^2}\Delta\frac{dg}{de}-\frac{2q}{e}$;
substituting it into Eq.~(\ref{dede1}), we obtain
\begin{eqnarray}\label{dede2}
\frac{dE}{de}=\omega\frac{dQ}{de}+
\int\left(\frac{1}{e^{2}}g\Delta\frac{dg}{de}-\frac{2}{e}gq-
\frac{1}{e^3}\partial_{i}g\partial_{i}g+\frac{1}{e^2}\partial_{i}g\partial_{i}\frac{dg}{de}\right)d^3x.
\end{eqnarray}
Substituting Eq.~(\ref{eqg1}) into (\ref{dede2}) and performing
integrations by parts in the resulting integral (since
$g|_{r\to\infty}\sim\frac{Q}{r}$ is assumed for gauged Q-balls and
consequently
$\frac{dg}{de}\bigl|_{r\to\infty}\sim\frac{dQ/de}{r}$, the
corresponding surface term vanishes again), we get
\begin{eqnarray}\label{dde}
\frac{dE}{de}=\omega\frac{dQ}{de}+\frac{1}{e^{3}}
\int\partial_{i}g\partial_{i}gd^3x.
\end{eqnarray}
Since $e$ is supposed to be nonnegative and $g\sim e^{2}$ for
$e\to 0$, we have
\begin{eqnarray}\label{dde1}
\frac{dE}{de}\ge\omega\frac{dQ}{de}.
\end{eqnarray}
Now we integrate Eq.~(\ref{dde1}) in $e$ from $e=0$ to $e=e_{x}$
and obtain
\begin{eqnarray}
E(Q(e_{x}),e_{x})-\omega Q(e_{x})>E(Q(0),0)-\omega Q(0).
\end{eqnarray}
$Q(0)$ and $E(Q(0),0)$ stand for the nongauged Q-ball, for which
$E(Q(0),0)-\omega Q(0)>0$ holds for any $\omega$, leading to
\begin{eqnarray}
E(Q(e_{x}),e_{x})-\omega Q(e_{x})>0.
\end{eqnarray}
The latter means that $\frac{E(Q_{min})}{Q_{min}}>\omega_{min}$
for such a gauged Q-ball, which implies that, according to the
reasonings presented above, the function $E_{aux}(Q)$ can be
constructed and the stability against fission for Q-balls
corresponding to the $\frac{d^{2}E}{dQ^{2}}<0$ branch of the
$E(Q)$ dependence can be established.

It is clear that the gauged Q-ball with $\frac{d^{2}E}{dQ^{2}}<0$
can not decay into a Q-ball from the same branch with
$\frac{d^{2}E}{dQ^{2}}<0$ and an anti-Q-ball (i.e., a Q-ball with
$\omega<0$ and $Q<0$). Indeed, for a gauged Q-ball we have
\begin{eqnarray}\nonumber
E(Q_{2})=E(Q_{2}+Q_{1})-\int\limits_{Q_{2}}^{Q_{2}+Q_{1}}\frac{dE}{d\tilde
Q}d\tilde Q=
E(Q_{2}+Q_{1})-\int\limits_{Q_{2}}^{Q_{2}+Q_{1}}\omega(\tilde
Q)d\tilde Q\\
\nonumber<E(Q_{2}+Q_{1})-\int\limits_{Q_{2}}^{Q_{2}+Q_{1}}\omega(\tilde
Q)d\tilde Q+E(-Q_{1})<E(Q_{2}+Q_{1})+E(-Q_{1}),
\end{eqnarray}
where $Q_{1}>0$ and $Q_{2}>0$, which means that such a decay is
energetically forbidden.

\section{Gauged Q-balls with small back-reaction of the gauge field}
\subsection{Linearized equations of motion}
It seems that it is very difficult, or even impossible, to find a
model providing an exact analytic solution for a gauged Q-ball in
the general case. Meanwhile, if the back-reaction of the gauge
field is supposed to be small ($|g(r)|\ll\omega$,
$|f(r)-f_{0}(r)|\ll f_{0}(r)$, where $f_{0}(r)=f_{0}(r,\omega)$ is
a nongauged Q-ball solution in the case $e=0$), one can try to use
the linear approximation in $g(r)$ and $\varphi(r)=f(r)-f_{0}(r)$
above the nongauged background solution, which simplifies the
analysis. In this case, equations (\ref{eqg1}) and (\ref{eqg2})
can be reduced to the form
\begin{eqnarray}\label{lin1}
\Delta g-2e^2\omega f_{0}^2=0,\\ \label{lin2}
\Delta\varphi+\omega^2\varphi+2\omega
gf_{0}-\frac{1}{2}\frac{d^2V}{df^2}\biggl|_{f=f_{0}}\varphi=0,
\end{eqnarray}
where $f_{0}$ is defined as a solution to the equation
\begin{eqnarray}\label{backgrsol}
\omega^2f_{0}+\Delta
f_{0}-\frac{1}{2}\frac{dV}{df}\biggl|_{f=f_{0}}=0
\end{eqnarray}
and the condition $|\varphi(r)|\ll f_{0}(r)$ is supposed to hold
for any $r$.\footnote{In fact, this condition is too stringent
and, as we will see later, can be relaxed.} To our knowledge, for
the first time, the system of equations (\ref{lin1}) and
(\ref{lin2}) with (\ref{backgrsol}) was analyzed in paper
\cite{Rosen}, in which the coupling constant $e$ was assumed to be
small. Note that here we do not put any restrictions on the
possible values of $e$; we only assume that the fields $g$ and
$\varphi$ are suppressed by the small factor proportional to
$e^2$. Although this factor is proportional to $e^{2}$, it does
not coincide with $e^{2}$ in the general case. This implies that
the linearized theory above the background solution $f_{0}$,
described by (\ref{lin1}) and (\ref{lin2}), can not be used when
only $e^{2}\ll 1$ holds --- the fields $g$ and $\varphi$ should
also remain small compared to $\omega$ and $f_{0}$ respectively.
We will discuss this issue in detail later; see also a simple
justification of this fact in Appendix~B.

\subsection{The charge and the energy of gauged Q-balls}
Linearizing the charge (\ref{chargedef}) and the energy
(\ref{energydef}) with respect to the background solution
$f_{0}(r)$, performing integration by parts and using the
linearized equations of motion, we arrive at
\begin{equation}\label{charge}
Q=Q_{0}+\triangle Q=Q_{0}+4\pi\int\limits_{0}^{\infty}dr
r^2(2gf_{0}^2+4\omega f_{0}\varphi),
\end{equation}
\begin{equation}\label{energy}
E=E_{0}+\triangle E=E_{0}+4\pi\omega\int\limits_{0}^{\infty}dr
r^2(gf_{0}^2+4\omega f_{0}\varphi),
\end{equation}
where $Q_{0}$ and $E_{0}$ are defined by
Eqs.~(\ref{chargedef}),~(\ref{energydef}) with the background
solution $f_{0}(r)$ for the scalar field and with $g\equiv 0$.

Now, let us calculate $\triangle Q$ and $\triangle E$. To this end
let us take equation (\ref{backgrsol}) and differentiate it with
respect to $\omega$. We get
\begin{eqnarray}\label{backgrsol2}
2\omega f_{0}+\omega^{2}\frac{df_{0}}{d\omega}+\Delta
\frac{df_{0}}{d\omega}-\frac{1}{2}\frac{d^{2}V}{df^{2}}\biggl|_{f=f_{0}}\frac{df_{0}}{d\omega}=0.
\end{eqnarray}
Now we take equation (\ref{lin2}), multiply it by
$\frac{df_{0}}{d\omega}$, integrate over the spatial volume and
perform integration by parts in the term containing $\Delta$. We
get
\begin{eqnarray}\label{integralpert}
\int\left(\varphi\left(\Delta\frac{df_{0}}{d\omega}+\omega^2\frac{df_{0}}{d\omega}
-\frac{1}{2}\frac{d^2V}{df^2}\biggl|_{f=f_{0}}\frac{df_{0}}{d\omega}\right)+2\omega
gf_{0}\frac{df_{0}}{d\omega}\right)d^{3}x=0.
\end{eqnarray}
Substituting (\ref{backgrsol2}) into (\ref{integralpert}), we
arrive at
\begin{eqnarray}\label{integralpert2}
\omega\int\left(gf_{0}\frac{df_{0}}{d\omega}-\varphi
f_{0}\right)d^{3}x=0.
\end{eqnarray}
Now, let us consider the charge (\ref{charge}). According to
(\ref{integralpert2})
\begin{equation}\label{charge2}
\triangle Q=4\pi\int\limits_{0}^{\infty}dr r^2(2gf_{0}^2+4\omega
f_{0}\varphi)=4\pi\int\limits_{0}^{\infty}dr
r^2\left(2gf_{0}^2+4\omega gf_{0}\frac{df_{0}}{d\omega}\right)=
4\pi\int\limits_{0}^{\infty}dr r^2g\frac{dq}{d\omega},
\end{equation}
where $q$ is defined as $q=2\omega f_{0}^2$. The last integral can
be transformed as
\begin{eqnarray}\nonumber
\int d^{3}x g\frac{dq}{d\omega}&=&\frac{d}{d\omega}\int
d^{3}xgq-\int d^{3}x q\frac{dg}{d\omega}=\frac{d}{d\omega}\int
d^{3}xgq-\int d^{3}x
\frac{1}{e^{2}}\Delta g\frac{dg}{d\omega}\\
\nonumber &=&\frac{d}{d\omega}\int d^{3}xgq-\int d^{3}x
\frac{1}{e^{2}}g\Delta\frac{dg}{d\omega}=\frac{d}{d\omega}\int
d^{3}xgq-\int d^{3}x
g\frac{dq}{d\omega},
\end{eqnarray}
where we have used (\ref{lin1}) and the relation $\Delta
\frac{dg}{d\omega}=e^{2}\frac{dq}{d\omega}$, which follows from
(\ref{lin1}). Thus, we get
\begin{equation}\label{forcharge2}
\int d^{3}x g\frac{dq}{d\omega}=\frac{1}{2}\frac{d}{d\omega}\int
d^{3}xgq.
\end{equation}
Let us define
\begin{equation}\label{Idef}
I=\frac{1}{2}\int d^{3}xgq.
\end{equation}
Then, from (\ref{charge2}) and (\ref{forcharge2}), we get
\begin{eqnarray}\label{chargeI}
\triangle Q=\frac{dI}{d\omega}.
\end{eqnarray}
Now it is easy to show that
\begin{eqnarray}\label{energyI}
\triangle E=\omega\triangle Q-I=\omega\frac{dI}{d\omega}-I.
\end{eqnarray}
With the help of (\ref{lin1}) the integral $I$ can be expressed in
the form
\begin{equation}\label{IelectrE}
I=-\frac{1}{2e^{2}}\int d^{3}x\partial_{i}g\partial_{i}g,
\end{equation}
which is nothing but the energy of the gauge field taken with the
minus sign (see (\ref{energydef})).

Relations (\ref{chargeI}) and (\ref{energyI}) allow us to check
explicitly the validity of (\ref{dEdQgauged}) in the linearized
theory. Indeed,
\begin{eqnarray}
\frac{d(E_{0}+\triangle E)}{d(Q_{0}+\triangle
Q)}=\frac{\frac{d(E_{0}+\triangle
E)}{d\omega}}{\frac{d(Q_{0}+\triangle
Q)}{d\omega}}=\frac{\omega\frac{dQ_{0}}{d\omega}+\omega\frac{d^{2}I}{d\omega^{2}}}{\frac{dQ_{0}}{d\omega}+\frac{d^{2}I}{d\omega^{2}}}=\omega.
\end{eqnarray}

Now, we turn to the calculation of the integral $I$. First, we
take equation (\ref{lin1}). Given a background solution $f_{0}$,
the spherically symmetric solution to (\ref{lin1}) such that
$g|_{r\to\infty}\to 0$, $\frac{dg}{dr}\bigl|_{r=0}=0$ takes the
form \cite{Rosen}
\begin{eqnarray}\label{Rosensol}
g=g(r)=-e^{2}\int\limits_{r}^{\infty}q(y)ydy-e^{2}\frac{1}{r}\int\limits_{0}^{r}q(y)y^{2}dy.
\end{eqnarray}
Substituting it into (\ref{Idef}), we get
\begin{equation}
I=-2\pi
e^{2}\left(\int\limits_{0}^{\infty}q(r)r^{2}\int\limits_{r}^{\infty}q(y)ydy\,dr+
\int\limits_{0}^{\infty}q(r)r\int\limits_{0}^{r}q(y)y^{2}dy\,dr\right).
\end{equation}
By performing integration by parts, it is easy to show that
\begin{equation}\label{equalintegrals}
\int\limits_{0}^{\infty}q(r)r^{2}\int\limits_{r}^{\infty}q(y)ydy\,dr=
\int\limits_{0}^{\infty}q(r)r\int\limits_{0}^{r}q(y)y^{2}dy\,dr.
\end{equation}
Thus, we arrive at
\begin{equation}\label{Ifinal}
\frac{I}{4\pi}=-4e^{2}\omega^{2}
\int\limits_{0}^{\infty}f_{0}^{2}(r)r\int\limits_{0}^{r}f_{0}^{2}(y)y^{2}dy\,dr,
\end{equation}
where we have used $q=2\omega f_{0}^2$. Equivalently,
(\ref{Ifinal}) can be rewritten as
\begin{equation}
\frac{I}{4\pi}=-2e^{2}\omega^{2}
\int\limits_{0}^{\infty}\left(\int\limits_{0}^{r}f_{0}^{2}(y)y^{2}dy\right)^{2}\frac{1}{r^{2}}\,dr.
\end{equation}

From (\ref{chargeI}), (\ref{energyI}) and (\ref{Ifinal}) we see
that the charge and the energy of a gauged Q-ball through the
terms linear in $e^{2}$ can be calculated using only the
background solution $f_{0}$ for the nongauged Q-ball. The
corresponding formulas look like
\begin{align}
\label{mainresultlin0} Q(\omega)&=Q_{0}(\omega)+\frac{dI(\omega)}{d\omega},\\
\label{mainresultlin}
E(\omega)&=E_{0}(\omega)+\omega\frac{dI(\omega)}{d\omega}-I(\omega),\\
\label{mainresultlin1} I(\omega)&=-16\,\pi e^{2}\omega^{2}
\int\limits_{0}^{\infty}f_{0}^{2}(r,\omega)r\int\limits_{0}^{r}f_{0}^{2}(y,\omega)y^{2}dy\,dr
\end{align}
where $Q_{0}(\omega)$ and $E_{0}(\omega)$ are the charge and the
energy of the nongauged Q-ball, respectively. Thus, to examine the
main properties of gauged Q-balls in a theory with a small
parameter (proportional to $e^2$) standing for the back-reaction
of the gauge field, it is not necessary to solve explicitly the
linearized differential equation (\ref{lin2}), which is a rather
complicated task and can be made only numerically in the general
case. Instead of this, one can simply take the corresponding
nongauged background solution $f_{0}(r,\omega)$, evaluate the
double integral in (\ref{Ifinal}) (numerically, in the general
case) to get the function $I(\omega)$, and calculate the
corresponding $E(Q)$ dependence.\footnote{As we will see below, in
most cases the energy of a gauged Q-ball at a given charge can be
calculated using the formula that is even simpler than those in
(\ref{mainresultlin0}) and (\ref{mainresultlin}).} We remind the
reader that for obtaining
(\ref{mainresultlin0})--(\ref{mainresultlin1}) we have used only
the supposition that $\varphi$ and $g$ are {\em exact} solutions
to linearized equations of motion; the restriction $e^{2}\ll 1$
has not been used.

\subsection{Validity criteria for the linear approximation}
It is clear that in a nonlinear theory the linear approximation
above a background solution is valid if the corrections are much
smaller than the background solution itself. In our case this
suggests that the relations
\begin{eqnarray}\label{vallin1}
|g(r)|\ll\omega,\\ \label{vallin2} |\varphi(r)|\ll f_{0}(r)
\end{eqnarray}
should be fulfilled for any $r$. We start with the first relation.
Eq.~(\ref{Rosensol}) implies that $\frac{dg}{dr}\ge 0$ for any $r$
and $g|_{r\to\infty}\to 0$ (of course, if the corresponding
integrals in (\ref{Rosensol}) converge, which is exactly our
case). This means that $|g(r)|\le |g(0)|$ for any $r$. The value
of $g(0)$ can be easily obtained from (\ref{Rosensol}) and takes
the form
\begin{eqnarray}\label{gtoomegadef}
g(0)=-2e^{2}\omega\int\limits_{0}^{\infty}f_{0}^{2}(y)ydy.
\end{eqnarray}
In this case (\ref{vallin1}) can be rewritten as
\begin{equation}\label{vallin3}
2e^{2}\int\limits_{0}^{\infty}f_{0}^{2}(y)ydy\ll 1.
\end{equation}
This inequality implies that the natural small parameter of the
theory is not simply $e^{2}$. Indeed, in principle it is possible
that even for a very small value of $e^{2}$ the integral
$\int\limits_{0}^{\infty}f_{0}^{2}(y)ydy$ is large enough and
inequality (\ref{vallin3}) is not fulfilled and vice versa. As we
will see below, the parameter $\frac{|g(0)|}{\omega}$ plays an
important role in the estimation of the small parameter of the
theory.

Now we turn to equation (\ref{vallin2}). As will be shown below by
particular examples, it is quite possible that
$\frac{|\varphi(r)|}{f_{0}(r)}$ grows with $r$ (this happens in
both models, which will be studied below; see also Appendix~C, in
which it is shown explicitly for a certain wide class of the
scalar field potentials). Formally, the linear approximation
breaks down at large $r$ in such a case. So, there arises a
questions: is it possible to use formulas
(\ref{mainresultlin0})--(\ref{mainresultlin1}), which were
obtained in the linear approximation, when it breaks down, though
at large $r$? The answer is yes, and below we will justify why it
is so.

To start with, let us suppose that there exists an exact solution
$f(r)$ to equations (\ref{eqg1}), (\ref{eqg2}) for a given
$\omega$, as well as a solution $f_{0}$ to Eq.~(\ref{backgrsol})
with the same $\omega$ as the one in $f$. Now, we take
(\ref{chargedef}) and (\ref{energydef}) and write the {\em exact}
equations
\begin{eqnarray}\nonumber
\triangle Q&=&Q-Q_{0}=\int d^{3}x\left(2gf_{0}^{2}+4\omega
(f-f_{0})f_{0}\right)\\ \label{deltaQexact}&+&\int
d^{3}x\left(2(\omega+g)(f-f_{0})^{2}+4g(f-f_{0}) f_{0}\right),\\
\nonumber\triangle E&=&E-E_{0}=\int d^{3}x\left(\omega
gf_{0}^{2}+4\omega^{2}(f-f_{0})f_{0}\right)+\int
d^{3}x\left(V(f)-V(f_{0})-(f-f_{0})\frac{dV}{df}\biggl|_{f=f_{0}}\right)
\\&+&\int d^{3}x\left(\omega^{2}(f-f_{0})^{2}+2\omega g(f-f_{0})
f_{0}+\omega
g(f-f_{0})^{2}+\partial_{i}(f-f_{0})\partial_{i}(f-f_{0})\right)\label{deltaEexact},
\end{eqnarray}
where we have used equations (\ref{eqg1}) and (\ref{backgrsol})
while performing integrations by parts. The functions $f_{0}$ and
$f$ are supposed to fall off rapidly at large $r$, so that
$f-f_{0}$ also falls off rapidly. Now let us assume that in the
inner region $r\le\hat R$, from which the main contribution to
$Q_{0}$ and $E_{0}$ comes, $|g|\ll\omega$ and
$|\varphi|=|f-f_{0}|\ll f_{0}$. The radius $\hat R$ can be defined
as
\begin{equation}\label{defRhat}
\int\limits_{\hat
R}^{\infty}f_{0}^{2}r^{2}dr=\epsilon\int\limits_{0}^{\hat
R}f_{0}^{2}r^{2}dr
\end{equation}
with $\epsilon\ll 1$ and $\hat R=\hat R(\omega)$. In this case, in
the inner region $r\le\hat R$ the last integrals in the rhs of
(\ref{deltaQexact}) and (\ref{deltaEexact}) can be neglected in
comparison with those containing only the linear terms in $g$ and
$\varphi$, whereas the second integral in the rhs of
(\ref{deltaEexact}) is equal to zero in this approximation. The
outer region $r>\hat R$ is supposed to be chosen such that the
fields $f$ and $f_{0}$ have very small absolute values inside it;
see (\ref{defRhat}). In this case, even though $\varphi$ can be of
the order of $f_{0}$ or larger, due to negligibly small {\em
absolute} values of the fields $f$ and $f_{0}$, we have (of
course, we suppose that $V(f)$ and $V(f_{0})$ are also negligibly
small in this area)
$$
\triangle Q_{\textrm{outer}}\ll \triangle Q_{\textrm{inner}},
$$
$$
\triangle E_{\textrm{outer}}\ll \triangle E_{\textrm{inner}},
$$
which leads to
\begin{eqnarray}\label{chargeapprox}
\triangle Q&\approx &4\pi\int\limits_{0}^{\hat R}
\left(2gf_{0}^{2}+4\omega\varphi f_{0}\right)r^{2}dr\approx
4\pi\int\limits_{0}^{\infty} \left(2gf_{0}^{2}+4\omega\varphi
f_{0}\right)r^{2}dr, \\ \label{energyapprox}\triangle E&\approx
&4\pi\omega\int\limits_{0}^{\hat R}\left(gf_{0}^{2}+4\omega
\varphi f_{0}\right)r^{2}dr\approx
4\pi\omega\int\limits_{0}^{\infty} \left(gf_{0}^{2}+4\omega
\varphi f_{0}\right)r^{2}dr
\end{eqnarray}
with a good accuracy. Of course, equations
(\ref{mainresultlin0})--(\ref{mainresultlin1}) are valid with the
same accuracy if the linearized theory works only in the inner
region. This is enough for all practical purposes.

The problem is how to check that the condition $|\varphi(r)|\ll
f_{0}(r)$ is valid in the inner region and the contribution of the
outer region is negligibly small. The only fully consistent way to
do it is to solve Eq.~(\ref{lin2}) with a particular
$f_{0}(r,\omega)$. For a fixed $\omega$ this can be done
numerically in the general case, the situation is more complicated
if one has to analyze a rather wide range of $\omega$: a search
for solutions to Eq.~(\ref{lin2}) and the subsequent calculation
of $\frac{|\varphi(r)|}{f_{0}(r)}$ for $r<\hat R$ may take quite a
long time. But we think that at least an estimate of the maximal
value of $\frac{|\varphi(r)|}{f_{0}(r)}$ for $r<\hat R$ can be
made without solving equation (\ref{lin2}).

Below we will propose two parameters, which can be useful for such
an estimate. To find the first parameter, we notice that if
$|\varphi(r)|\ll f_{0}(r)$, then
\begin{equation}\label{estimate1}
\biggl|\int\limits_{0}^{\hat R}\varphi f_{0}r^{2}dr\biggr|\ll
\biggl|\int\limits_{0}^{\hat R}f_{0}^{2}r^{2}dr\biggr|.
\end{equation}
The opposite is not correct. Indeed, if $\varphi$ changes its sign
at some $r>0$ (this is exactly the situation realized in Model~2,
which will be presented below), then it is possible that the
integral in the lhs of (\ref{estimate1}) is equal to zero for
nonzero $\varphi$, which does not provide any estimate. Meanwhile,
inequality (\ref{estimate1}) may be useful taken together with
other parameters, which will be discussed later. For $r>\hat R$,
the absolute values of the field $\varphi$ are negligibly small in
the outer region, so we can rewrite inequality (\ref{estimate1})
as
\begin{equation}\label{varf}
\biggl|\int\limits_{0}^{\infty}\varphi f_{0}r^{2}dr\biggr|\ll
\biggl|\int\limits_{0}^{\infty}f_{0}^{2}r^{2}dr\biggr|.
\end{equation}
In other words, the fulfillment of the latter inequality implies
that the contribution of the fields from the outer region, where
$|\varphi|$ can be of the order of $f_{0}$ and larger, is
negligibly small in comparison with the main contribution of the
inner region. Now, multiplying (\ref{varf}) by $4\pi\omega^2$ and
using (\ref{chargeapprox}) and (\ref{energyapprox}) (or
(\ref{charge}) and (\ref{energy})), we can rewrite (\ref{varf}) as
\begin{equation}\label{vallin4}
\frac{|2\triangle E-\omega\triangle Q|}{2\omega Q_{0}}\ll 1.
\end{equation}
It can be rewritten in the explicit form as
\begin{equation}
\biggl|{\frac{1}{4\omega}\frac{dI}{d\omega}-\frac{I}{2\omega^{2}}}\biggr|\ll{4\pi\int\limits_{0}^{\infty}f_{0}^{2}r^{2}dr}.
\end{equation}

Now we turn to the second parameter. To find it we take the
inhomogeneous equation (\ref{lin2}) and rewrite it as
\begin{eqnarray}
\label{lin2approx0}
\Delta\varphi+\omega^2\varphi-\frac{1}{2}\frac{d^2V}{df^2}\biggl|_{f=f_{0}}\varphi=-2\omega
g(r)f_{0}(r)\le -2\omega g(0)f_{0}(r),
\end{eqnarray}
where we have used the fact that $g(r)<0$ and $|g(r)|\le|g(0)|$
for any $r$. Eq.~(\ref{lin2approx0}) suggests that, at least for
an estimation of $\varphi$, one can consider the simplified
equation
\begin{eqnarray}
\label{lin2approx1}
\Delta\hat\varphi+\omega^2\hat\varphi-\frac{1}{2}\frac{d^2V}{df^2}\biggl|_{f=f_{0}}\hat\varphi+2\omega
g(0)f_{0}(r)=0,
\end{eqnarray}
instead of (\ref{lin2}). We think that the difference between
$\varphi$ and $\hat\varphi$ should be of the order of $\varphi$,
which is not critical for the estimate. But according to
Eq.~(\ref{backgrsol2}), Eq.~(\ref{lin2approx1}) can be solved
exactly
--- its solution has the form
\begin{equation}
\hat\varphi=g(0)\frac{df_{0}}{d\omega}.
\end{equation}
Thus, instead of $\frac{|\varphi(r)|}{f_{0}(r)}$ we can try to
estimate $\frac{|\hat\varphi(r)|}{f_{0}(r)}$, for which
\begin{equation}\label{parameter2}
\biggl|\frac{g(0)}{f_{0}(r)}\frac{df_{0}(r)}{d\omega}\biggr|\ll 1
\end{equation}
should hold. Note that the new parameter in (\ref{parameter2}) is
proportional to the first parameter $\frac{|g(0)|}{\omega}$. We
think that in order to get better estimates one should calculate
(\ref{parameter2}) at several different points of $r$ for a given
$\omega$.

The fulfillment of (\ref{vallin3}) together with (\ref{vallin4})
and (\ref{parameter2}) suggests, although it does not ensure, that
the linear approximation is valid for $g$ and $\varphi$ in the
inner region, whereas the outer region does not make any
significant contribution, and the linearized theory indeed can be
used for a description of gauged Q-ball.\footnote{Of course, the
breakdown of the linear approximation at large $r$ does not mean
that a solution to nonlinear equations (\ref{eqg1}) and
(\ref{eqg2}) does not exist
--- it simply means that the linear approximation does not
describe the Q-ball properly far away from its center.} Note that,
as will be shown below by an explicit example (Model 2), the
fulfillment of (\ref{vallin3}) does not imply the fulfillment of
(\ref{vallin4}) and (\ref{parameter2}) and vice versa. Thus, in
the general case, one should estimate {\em all} the parameters
presented above while analyzing the question about the
applicability of (\ref{mainresultlin0})--(\ref{mainresultlin1}).
It is possible simply to define the function
\begin{eqnarray}\label{expparam}
\alpha(\omega)=\underset{i}{\textrm{max}}\left\{\frac{|g(0)|}{\omega},\frac{|2\triangle
E-\omega\triangle Q|}{2\omega
Q_{0}},\biggl|\frac{g(0)}{f_{0}(r_{i})}\frac{df_{0}(r_{i})}{d\omega}\biggr|\right\}
\end{eqnarray}
and consider it as the natural small parameter depending on
$\omega$, for which
\begin{eqnarray}\label{expparam1}
\alpha(\omega)\ll 1
\end{eqnarray}
should hold. We emphasize that for calculating $\alpha(\omega)$
only the background solution $f_{0}(r,\omega)$ is necessary.
Because of the dependence of the parameter $\alpha(\omega)$ on
$\omega$, one can say that it ``runs'' with $\omega$. It is
obvious that $\alpha(\omega)\sim e^{2}$, but, as we will see
below, the smallness of $e^{2}$ does not guarantee the fulfillment
of (\ref{expparam1}).

\subsection{Comparison of gauged and nongauged Q-balls}
Now let us compare some properties of gauged (obtained in the
linear approximation in $\alpha(\omega)$) and nongauged Q-balls.
We start with comparing the energies of Q-balls at a given charge
$Q$. For a gauged Q-ball we have $Q=Q_{0}(\omega_{1})+\triangle
Q(\omega_{1})$, whereas for a nongauged Q-ball
$Q=Q_{0}(\omega_{2})$. From $Q_{0}(\omega_{1})+\triangle
Q(\omega_{1})=Q_{0}(\omega_{2})$, in the linear approximation in
$\alpha(\omega)$, we obtain
\begin{equation}\label{deltaomega}
\triangle
Q(\omega_{1})=(\omega_{2}-\omega_{1})\frac{dQ_{0}}{d\omega}\biggl|_{\omega=\omega_{1}}.
\end{equation}
Now let us compare the energies of gauged and nongauged Q-balls
with the same charges. We get
\begin{eqnarray}\nonumber
&&E(\omega_{1})-E_{0}(\omega_{2})=E_{0}(\omega_{1})+\triangle
E(\omega_{1})-E_{0}(\omega_{2})\approx\triangle E(\omega_{1})-(\omega_{2}-\omega_{1})\frac{dE_{0}}{d\omega}\biggl|_{\omega=\omega_{1}}\\
&&=\triangle
E(\omega_{1})-\frac{\frac{dE_{0}}{d\omega}\bigl|_{\omega=\omega_{1}}}{\frac{dQ_{0}}{d\omega}\bigl|_{\omega=\omega_{1}}}\triangle
Q(\omega_{1})=\triangle E(\omega_{1})-\omega_{1}\triangle
Q(\omega_{1}),\label{E-E0}
\end{eqnarray}
where we have used Eq.~(\ref{deltaomega}) and the relation
$\frac{dE_{0}}{dQ_{0}}=\omega$. But according to (\ref{chargeI}),
(\ref{energyI}), and (\ref{IelectrE}),
\begin{eqnarray}\label{positive}
\triangle E(\omega_{1})-\omega_{1}\triangle
Q(\omega_{1})=-I(\omega_{1})=\frac{1}{2e^{2}}\int
d^{3}x\partial_{i}g\partial_{i}g,
\end{eqnarray}
which is always positive for $g\not\equiv 0$. Thus, for any charge
$Q>0$ the energy of a gauged Q-ball is larger than the energy of
the corresponding nongauged Q-ball with the same charge that is,
of course, the expected result. It is interesting to note that
(\ref{positive}) also follows from (\ref{dde}). Indeed, $\triangle
Q\sim e^{2}$ and $\triangle E\sim e^{2}$, leading to
$\frac{dQ}{de}=\frac{2\triangle Q}{e}$ and
$\frac{dE}{de}=\frac{2\triangle E}{e}$. Substituting the latter
relations into (\ref{dde}), we get (\ref{positive}).

Now, we turn to examining another property of Q-balls --- cusps on
the $E(Q)$ diagrams. Such cusps, which are a consequence of the
existence of (locally) minimal or/and (locally) maximal charges,
exist on $E(Q)$ diagrams in many models of nongauged Q-balls. The
origin of the cusps is the following: for a (locally) minimal or a
(locally) maximal charge, we have
$\frac{dQ}{d\omega}\bigl|_{\omega=\hat\omega}=0$ with
$\hat\omega>0$, whereas from
$\frac{dE}{d\omega}=\omega\frac{dQ}{d\omega}$, it follows that
$\frac{dE}{d\omega}\bigl|_{\omega=\hat\omega}=0$, which leads to
the appearance of a cusp at the point $Q_{m}=Q(\hat\omega)$. Of
course, analogous cusps are expected in the gauged case also. For
example, one can recall the model of \cite{Lee:1991bn}, in which
the function $E(Q)$ was drawn with the help of gauged
non-topological soliton solutions, which were obtained by solving
numerically the corresponding exact equations of motion, although
for rather small values of the expansion parameter (the cusp is
clearly seen on the $E(Q)$ diagram presented in
\cite{Lee:1991bn}). Below, we will obtain relations between the
charges, corresponding to the cusps, in the gauged and nongauged
cases.

The position of a cusp in the gauged case is defined by
$\frac{d(Q_{0}+\triangle
Q)}{d\omega}\bigl|_{\omega=\hat\omega_{1}}=0$, whereas for the
nongauged case it is defined by
$\frac{dQ_{0}}{d\omega}\bigl|_{\omega=\hat\omega_{2}}=0$. The
difference between the charges in the linear order in
$\alpha(\omega)$ is
\begin{eqnarray}\nonumber
Q(\hat\omega_{1})-Q_{0}(\hat\omega_{2})=Q_{0}(\hat\omega_{1})+\triangle
Q(\hat\omega_{1})-Q_{0}(\hat\omega_{2})\\ \label{cusppos}\approx
\triangle
Q(\hat\omega_{1})+(\hat\omega_{1}-\hat\omega_{2})\frac{dQ_{0}}{d\omega}\bigl|_{\omega=\hat\omega_{2}}=\triangle
Q(\hat\omega_{1})\approx \triangle Q(\hat\omega_{2}).
\end{eqnarray}
We see that in the linear approximation in $\alpha(\omega)$ the
difference between the charges corresponding to the cusps in the
gauged and nongauged cases is defined by the value of $\triangle
Q$ at $\omega$ corresponding to the cusp in the nongauged case. As
we will see below using explicit examples, the difference can be
positive, negative, or even zero.

One makes an interesting observation from equations (\ref{E-E0}),
(\ref{positive}), and (\ref{cusppos}). Since $I(\omega_{1})\approx
I(\omega_{2})$ in the linear order in $\alpha(\omega)$, for
$\omega_{2}$ which is not very close to $\hat\omega_{2}$, one has
$E(\omega_{1})=E_{0}(\omega_{2})-I(\omega_{2})$. Suppose that we
have a nongauged Q-ball with the charge $Q_{x}$ and the energy
$E_{0}(Q_{x})$. Then, the energy of the corresponding gauged
Q-ball with the same {\em charge} $Q_{x}$ (not with the same
$\omega$) is simply
\begin{equation}\label{EgE0-main}
E(Q_{x})=E_{0}(Q_{x})-I(\omega)|_{\omega=Q_{0}^{-1}(Q_{x})},
\end{equation}
where $-I(\omega)|_{\omega=Q_{0}^{-1}(Q_{x})}$ is just the energy
of the gauge field produced by the nonpointlike charge $Q_{x}$
(recall Eq.~(\ref{IelectrE})). Near the cusps, this formula must
be used very carefully: at first, it is necessary to check that
for a given charge $Q_{x}$ of the nongauged Q-ball the
corresponding gauged Q-ball really exists (see
Eq.~(\ref{cusppos})) and that (\ref{deltaomega}) results in
$\omega_{2}-\omega_{1}\sim\alpha(\omega_{2})$. If it is not so,
one should use (\ref{mainresultlin0}) and (\ref{mainresultlin})
instead of (\ref{EgE0-main}). But it is clear that
(\ref{EgE0-main}) can be used for most values of the Q-ball
charge.

\subsection{Explicit examples of gauged Q-balls}
In the general case, a straightforward numerical evaluation of the
function $I(\omega)$ for a given background solution
$f_{0}(r,\omega)$ may take quite a long time. So, to illustrate
how the general results, presented above, can be used for
calculations, we choose two models with very simple background
Q-ball solutions $f_{0}(r,\omega)$. The simplicity of the
background solutions allows us not only to find the function
$I(\omega)$ analytically in both cases but also to obtain {\em
exact} analytic solutions to the system of linearized equations
(\ref{lin1}), (\ref{lin2}).

\subsubsection{Model 1} Let us consider the model proposed in
\cite{Rosen:1969ay} with the potential (in our notations)
\begin{equation}
V(\phi^*\phi)=-\mu^2\phi^*\phi\ln(\beta^2\phi^*\phi),
\label{potential1}
\end{equation}
where $\mu$ and $\beta$ are the model parameters. The spherically
symmetric background (nongauged) solution for the Q-ball in this
model takes the form
\begin{eqnarray}\label{backsolRosen}
f_{0}(r)=\mu\xi\textrm{e}^{-\frac{\omega^2}{2\mu^2}}\textrm{e}^{-\frac{\mu^2r^2}{2}},
\end{eqnarray}
where $0\le\omega<\infty$ and $\xi=\frac{\textrm{e}}{\beta\mu}$.
The charge and the energy of the Q-ball look like
\begin{eqnarray}\label{charge3DRosen}
Q_{0}&=&2\pi^{\frac{3}{2}}\xi^{2}\frac{\omega}{\mu}\,\textrm{e}^{-\frac{\omega^2}{\mu^2}},\\
\label{energy3DRosen}
E_{0}&=&2\pi^{\frac{3}{2}}\xi^{2}\mu\left(\frac{\omega^{2}}{\mu^{2}}+\frac{1}{2}\right)\textrm{e}^{-\frac{\omega^2}{\mu^2}}.
\end{eqnarray}
For additional details concerning nongauged Q-balls in the model
with potential (\ref{potential1}), see \cite{MarcVent}, in which
this model was thoroughly investigated.

\begin{figure}[!ht]
\begin{center}
\includegraphics[width=0.9\linewidth]{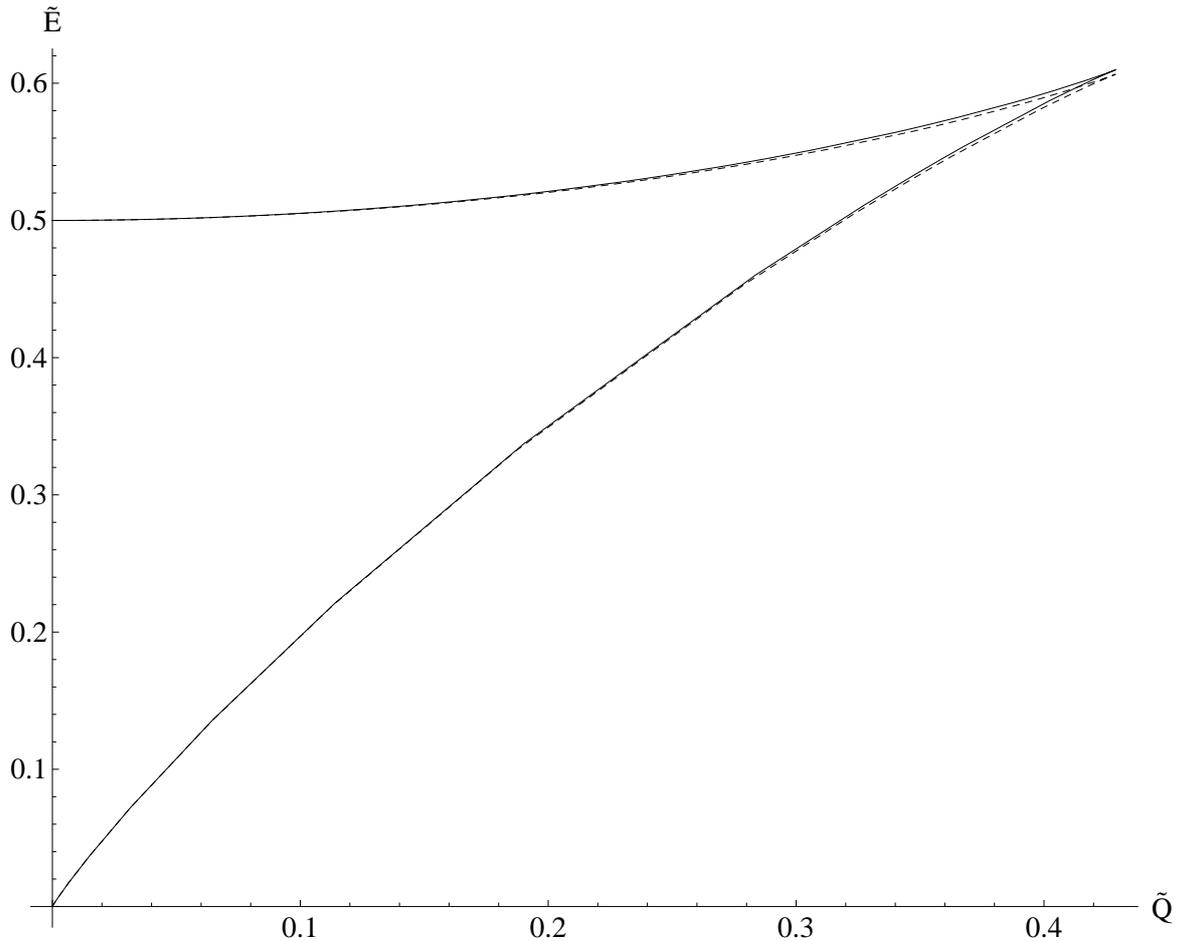}
\caption{$E(Q)$ for the gauged (solid line) and nongauged (dashed
line) cases. Here, $\alpha_{1}=0.05$ and $0\le\tilde\omega\le
10$.}\label{EQ_Rosen}
\end{center}
\end{figure}

The integral in (\ref{mainresultlin1}) can be easily calculated
analytically for the background solution defined by
(\ref{backsolRosen}). The result looks like
\begin{eqnarray}\label{IRosenmodel}
\frac{I}{4\pi}=-\mu
e^{2}\frac{\sqrt{\pi}}{4\sqrt{2}}\xi^{4}\left(\frac{\omega}{\mu}\right)^{2}\textrm{e}^{-\frac{2\,\omega^2}{\mu^2}}.
\end{eqnarray}
We see that (\ref{IRosenmodel}) has a very simple form. The
corrections $\triangle Q$ and $\triangle E$ can also be calculated
analytically, and for the charge and the energy of the gauged
Q-ball we get
\begin{eqnarray}\label{QdimensRosen}
Q&=&Q_{0}+\triangle Q=2\pi^{\frac{3}{2}}\xi^{2}\left(\tilde
Q_{0}+e^{2}\xi^{2}\triangle\tilde
Q\right)=2\pi^{\frac{3}{2}}\xi^{2}\tilde Q,\\
\label{EdimensRosen}E&=&E_{0}+\triangle
E=\mu\,2\pi^{\frac{3}{2}}\xi^{2}\left(\tilde
E_{0}+e^{2}\xi^{2}\triangle\tilde
E\right)=\mu\,2\pi^{\frac{3}{2}}\xi^{2}\tilde E
\end{eqnarray}
with
\begin{eqnarray}\label{Q0dimensRosen}
\tilde Q_{0}&=&\tilde\omega\textrm{e}^{-\tilde\omega^{2}},\\
\tilde
E_{0}&=&\left(\tilde\omega^2+\frac{1}{2}\right)\textrm{e}^{-\tilde\omega^{2}},\\
\label{deltaQRosen}\triangle\tilde Q&=&\left(\sqrt{2}\,\tilde\omega^3-\frac{\tilde\omega}{\sqrt{2}}\right)\textrm{e}^{-2\,\tilde\omega^{2}},\\
\label{deltaERosen}\triangle\tilde
E&=&\left(\sqrt{2}\,\tilde\omega^4-\frac{\tilde\omega^{2}}{2\sqrt{2}}\right)\textrm{e}^{-2\,\tilde\omega^{2}},
\end{eqnarray}
where $\tilde\omega=\frac{\omega}{\mu}$. We also define the
parameter
\begin{equation}\label{alpha1}
\alpha_{1}=e^2\xi^2,
\end{equation}
which will be used below. In Fig.~\ref{EQ_Rosen} one can see an
example of the $E(Q)$ diagram for the gauged Q-ball in this model.
This diagram was plotted using formulas (\ref{mainresultlin0}) and
(\ref{mainresultlin}). We see from Fig.~\ref{EQ_Rosen} that the
energy of the gauged Q-ball is larger than the energy of the
corresponding nongauged Q-ball with the same charge, as it was
shown in the previous section. One sees that there is a cusp on
the $E(Q)$ diagram for the gauged case as well as for the
nongauged case. There are maximal charges, which correspond to
these cusps. For the nongauged case the charge is maximal at
$\tilde\omega=\frac{1}{\sqrt{2}}$. From (\ref{deltaQRosen}) it
follows that $\triangle Q|_{\tilde\omega=\frac{1}{\sqrt{2}}}=0$,
which means that the values of the maximal charge in the gauged
and nongauged cases coincide through the linear order in
$\alpha_{1}$ (this also implies that one can use (\ref{EgE0-main})
instead of (\ref{mainresultlin0}) and (\ref{mainresultlin}) for
plotting the $E(Q)$ diagram, presented in Fig.~\ref{EQ_Rosen},
with the same accuracy). It follows from Fig.~\ref{EQ_Rosen} that
on the lower branch of the $E(Q)$ diagram
$\frac{d^{2}E}{dQ^{2}}<0$ and $E(0)=0$ (the latter corresponds to
$\omega\to\infty$), which means that gauged Q-balls from this
branch are stable against fission.

\begin{figure}[h]
\begin{center}
\includegraphics[width=0.9\linewidth]{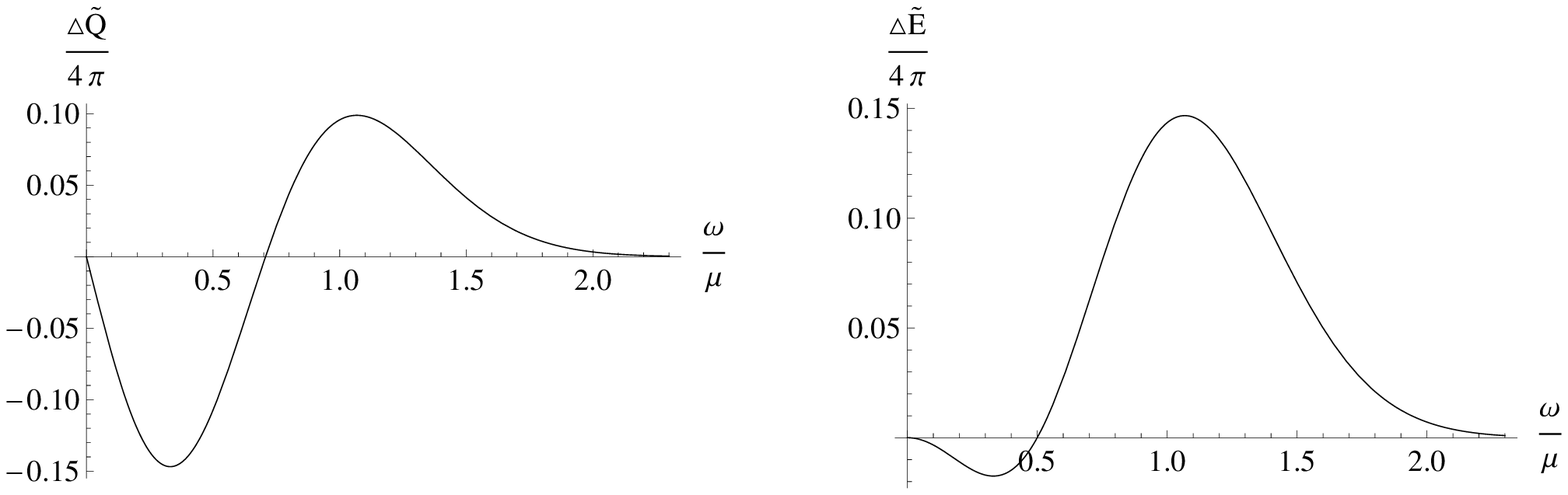}
\caption{$\triangle\tilde Q$ (left plot) and $\triangle\tilde E$
(right plot) for $0\le\tilde\omega\le 2.3$.}\label{EQ_Rosen-corr2}
\end{center}
\end{figure}

In Fig.~\ref{EQ_Rosen-corr2}, the plots of corrections
$\triangle\tilde Q$ and $\triangle\tilde E$ are presented. One
sees from these plots that $\triangle Q$ and $\triangle E$ can be
negative or positive for a given $\omega$ (although the energy of
gauged Q-ball is always larger than the energy of the
corresponding nongauged Q-ball with the same charge).

The parameter $\frac{|2\triangle E-\omega\triangle Q|}{2\omega
Q_{0}}$ in (\ref{vallin4}), which is necessary for checking the
applicability of linear approximation, can be obtained directly
from (\ref{Q0dimensRosen}), (\ref{deltaQRosen}), and
(\ref{deltaERosen}). It is not difficult to show that it can be
estimated as
\begin{eqnarray}\label{vallin4Rosen}
\frac{|2\triangle E-\omega\triangle Q|}{2\omega
Q_{0}}=\frac{\alpha_{1}}{\sqrt{2}}\,\tilde\omega^{2}\textrm{e}^{-\tilde\omega^{2}}\le
\frac{1}{\sqrt{2}\,\textrm{e}}\alpha_{1}.
\end{eqnarray}
As for the parameter $\frac{|g(0)|}{\omega}$ in (\ref{vallin3}),
it can also be calculated analytically for (\ref{backsolRosen})
and takes the form
\begin{equation}\label{stringentpar}
\frac{|g(0)|}{\omega}=\alpha_{1}\,\textrm{e}^{-\tilde\omega^{2}}\le
\alpha_{1}.
\end{equation}
And finally, the parameter (\ref{parameter2}) does not depend on
$r$ in this model and has the form
\begin{equation}\label{parameter2est1}
\biggl|\frac{g(0)}{f_{0}(r)}\frac{df_{0}(r)}{d\omega}\biggr|=
\alpha_{1}\tilde\omega^{2}\textrm{e}^{-\tilde\omega^{2}}\le\frac{1}{\textrm{e}}\alpha_{1}.
\end{equation}

We see that in the model under consideration all the parameters
$\frac{|g(0)|}{\omega}$, $\frac{|2\triangle E-\omega\triangle
Q|}{2\omega Q_{0}}$, and
$\left|\frac{g(0)}{f_{0}(r)}\frac{df_{0}(r)}{d\omega}\right|$ can
be easily estimated. The most stringent $\omega$-independent
restriction on $e\xi$ comes from (\ref{stringentpar}) and looks
very simple:
$$e^{2}\xi^{2}=\alpha_{1}\ll 1.$$
Note, that for large $\tilde\omega$ Eq.~(\ref{expparam}) gives
$$\alpha(\omega)=e^{2}\xi^{2}\tilde\omega^{2}\textrm{e}^{-\tilde\omega^{2}}.$$
It means that, with a fixed $e\xi$, the larger $\tilde\omega$ is,
the smaller the parameter $\alpha(\omega)$ is. In other words, the
larger $\tilde\omega$ is, the larger the maximal value of $e\xi$,
for which the linearized theory can be used with this
$\tilde\omega$, is. Nevertheless, one can consider $\alpha_{1}$ as
an $\omega$-independent small parameter for this model, which can
be useful in certain cases.\nopagebreak

The restriction $\alpha_{1}\ll 1$ clearly shows that the
fulfillment of $e^2\ll 1$ is not sufficient to ensure the validity
of the linear approximation. Indeed, even for a very small value
of $e^{2}$, the parameter $\xi$, which is defined by the
parameters of the scalar field potential, can be rather large to
make the use of the linear approximation impossible (this fact was
previously observed in \cite{Dzhunushaliev:2012zb}).

For completeness below we present the explicit solution for the
fields $g$ and $\varphi$ in this model. It satisfies the
conditions $\frac{dg}{dr}\bigl|_{r=0}=0$, $g|_{r\to\infty}=0$,
$\frac{d\varphi}{dr}\bigl|_{r=0}=0$, and $\varphi|_{r\to\infty}=0$
and can be factorized into terms containing $\omega$ and $r$. For
the first time, this exact, in the linear approximation, solution
was obtained in \cite{Dzhunushaliev:2012zb}, and in our notations,
it has the form
\begin{eqnarray}\label{linsolRosenbegin}
g(r)&=&\mu\alpha_{1}\,\Phi_{g}(\omega)F_{g}(r),\\
\varphi(r)&=&\mu\alpha_{1}\xi\,\Phi_{\varphi}(\omega)F_{\varphi}(r),
\end{eqnarray}
where
\begin{eqnarray}
\Phi_{g}(\omega)&=&\frac{\sqrt{\pi}}{2}\frac{\omega}{\mu}\,\textrm{e}^{-\frac{\omega^2}{\mu^2}},\\
F_{g}(r)&=&-\frac{1}{\mu\,r}\textrm{erf}(\mu\,r),\\
\Phi_{\varphi}(\omega)&=&\sqrt{\pi}\left(\frac{\omega}{\mu}\right)^{2}\textrm{e}^{-\frac{3\omega^2}{2\mu^2}},\\
F_{\varphi}(r)&=&\textrm{e}^{-\frac{3\mu^{2}r^{2}}{2}}\left(\frac{1}{4\sqrt{\pi
}}+\frac{1}{4}\,\textrm{e}^{\,\mu^{2}r^{2}}\left(\mu
r+\frac{1}{2\mu r}\right)\textrm{erf}(\mu
r)\right).\label{Fvarphi}
\end{eqnarray}
Here
$\textrm{erf}(z)=\frac{2}{\sqrt{\pi}}\int\limits_{0}^{z}\textrm{e}^{-t^{2}}dt$.

\begin{figure}[!h]
\begin{center}
\includegraphics[width=0.8\linewidth]{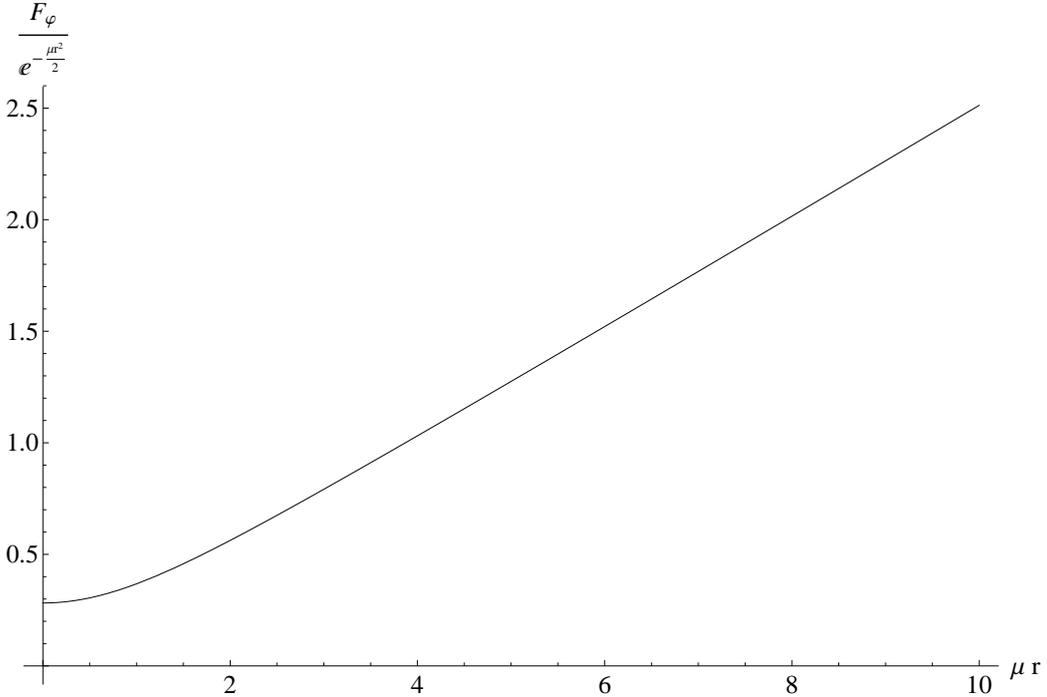}
\caption{The function
$\frac{F_{\varphi}(r)}{\textrm{e}^{-\frac{\mu^2r^2}{2}}}$.}\label{growth}
\end{center}
\end{figure}
The explicit solution, presented above, allows us to estimate how
$\frac{|\varphi(r)|}{f_{0}(r)}$ grows with $r$. In
Fig.~\ref{growth} the function
$\frac{F_{\varphi}(r)}{\textrm{e}^{-\frac{\mu^2r^2}{2}}}$ is
presented, clearly indicating the growth. One sees from this plot
that, for example, for a given $\omega$ the value of
$\frac{\varphi(r)}{f_{0}(r)}\bigl|_{\mu r=10}$ is approximately
five times larger than the value of
$\frac{\varphi(r)}{f_{0}(r)}\bigl|_{\mu r=2}$. Meanwhile, the
absolute values of the fields $\varphi$ and $f_{0}$ are
proportional to the factor of the order of $\textrm{e}^{-50}$ at
$\mu r=10$, which is extremely small. This confirms that if, with
an appropriate choice of $\alpha_{1}$, the linear approximation is
valid in the inner region of this gauged Q-ball (which can be
defined, for example, as $0\le\mu r\le 10$), formulas
(\ref{IRosenmodel})--(\ref{deltaERosen}) are also valid.

\subsubsection{Model 2}
Now we consider the model with a piecewise parabolic potential,
which was proposed in \cite{Rosen0} and thoroughly examined in
\cite{Gulamov:2013ema}.\footnote{Another model with a piecewise
parabolic potential (it was also proposed in \cite{Rosen0}),
admitting a rather simple solution, was discussed in detail in
\cite{Theodorakis:2000bz}. The model in \cite{Gulamov:2013ema}
provides a simpler solution (especially for $R(\omega)$, which,
contrary to the case of \cite{Theodorakis:2000bz}, has a very
simple analytic form (\ref{Rdef})), which appears to be more
useful for illustrative purposes and numerical analysis.} The
piecewise scalar field potential in this model has the form
\begin{equation}
V(\phi^*\phi)=M^2\phi^*\phi\,\theta\left(1-\frac{\phi^*\phi}{v^2}\right)+\left(m^2\phi^*\phi+v^2(M^2-m^2)\right)\theta\left(\frac{\phi^*\phi}{v^2}-1\right),
\label{potential2}
\end{equation}
where $M^2>0$, $M^2>m^2$, and $\theta$ is the Heaviside step
function with the convention $\theta(0)=\frac{1}{2}$. The
background solution for the Q-ball in this model takes the form
\begin{eqnarray}\label{backsol1}
f_{0}(r<R)=f_{0}^{<}(r)&=&v\frac{R\sin\left(\sqrt{\omega^{2}-m^{2}}\,r\right)}{r\sin\left(\sqrt{\omega^{2}-m^{2}}\,R\right)},\\
\label{backsol2} f_{0}(r>R)=f_{0}^{>}(r)&=&v\frac{R
\textrm{e}^{-\sqrt{M^2-\omega^2}r}}{r
\textrm{e}^{-\sqrt{M^2-\omega^2}R}},
\end{eqnarray}
where $R$ is defined as
\begin{equation}\label{Rdef}
R=R(\omega)=\frac{1}{\sqrt{\omega^{2}-m^{2}}}\left(\pi-\arctan\left(\frac{\sqrt{\omega^{2}-m^{2}}}{\sqrt{M^2-\omega^2}}\right)\right).
\end{equation}
The charge and the energy of the Q-ball look like
\begin{eqnarray}\label{charge3D}
Q_{0}&=&4\pi R^2\omega
v^2\left(\frac{(M^2-m^2)(R\sqrt{M^2-\omega^2}+1)}{(\omega^2-m^{2})\sqrt{M^2-\omega^2}}\right),\\
\label{energy3D} E_{0}&=&\omega Q_{0}+4\pi
\frac{R^{3}v^{2}(M^2-m^2)}{3}.
\end{eqnarray}
For additional details concerning nongauged Q-balls in the model
with potential (\ref{potential2}), see \cite{Gulamov:2013ema}.

As in the previous case, the integral in (\ref{mainresultlin1})
can be calculated analytically for the background solution defined
by (\ref{backsol1}) and (\ref{backsol2}). The result looks like
\begin{eqnarray}\nonumber
&\frac{I}{4\pi}=e^{2}\omega^{2}\left[a^{4}\left(\frac{\sin(2\sqrt{\omega^{2}-m^{2}}\,R)}{2\sqrt{\omega^{2}-m^{2}}\,}-R+
\frac{\textrm{Si}(2\sqrt{\omega^{2}-m^{2}}\,R)}{2\sqrt{\omega^{2}-m^{2}}\,}-\frac{\textrm{Si}(4\sqrt{\omega^{2}-m^{2}}\,R)}
{4\sqrt{\omega^{2}-m^{2}}\,}\right)\right.&\\
\nonumber&-\left.4b^{2}\left(a^{2}\left(\frac{R}{2}-\frac{\sin(2\sqrt{\omega^{2}-m^{2}}\,R)}{4\sqrt{\omega^{2}-m^{2}}\,}\right)+
\frac{b^{2}\textrm{e}^{-2\sqrt{M^2-\omega^2}\,R}}{2\sqrt{M^2-\omega^2}}\right)\textrm{E}_{1}(2\sqrt{M^2-\omega^2}\,R)\right.&\\
\label{Iourmodel}&+\left.\frac{2b^{4}}{\sqrt{M^2-\omega^2}}\textrm{E}_{1}(4\sqrt{M^2-\omega^2}\,R)\right],&
\end{eqnarray}
where
\begin{eqnarray}
\textrm{Si}(y)&=&\int\limits_{0}^{y}\frac{\sin(t)}{t}dt,\\
\textrm{E}_{1}(y)&=&\int\limits_{y}^{\infty}\frac{\textrm{e}^{-t}}{t}dt
\end{eqnarray}
and
\begin{eqnarray}\label{coefa}
a&=&a(\omega)=\frac{vR}{\sin\left(\sqrt{\omega^{2}-m^{2}}\,R\right)},\\
\label{coefb}b&=&b(\omega)=\frac{vR}{\textrm{e}^{-\sqrt{M^2-\omega^2}R}}.
\end{eqnarray}
We see that (\ref{Iourmodel}) has a much more complicated form
than the corresponding result for the previous model. In
principle, with the help of (\ref{Rdef}), (\ref{coefa}) and
(\ref{coefb}) the derivative $\frac{dI}{d\omega}$ can also be
calculated analytically, although we derived it numerically for
obtaining the $E(Q)$ dependence.

To perform numerical calculations, one should pass to
dimensionless variables. The most natural choice for the scale
parameter in this model is the mass parameter $M$. Thus, we choose
the dimensionless variables $\tilde\omega=\frac{\omega}{M}$ and
$\tilde r=Mr$. The background scalar field takes the form
\begin{eqnarray}\label{fdimens}
f_{0}(\omega,r)=v\tilde f_{0}(\tilde\omega,\tilde r).
\end{eqnarray}
It is not difficult to show that the charge and the energy can be
represented as
\begin{eqnarray}\label{Qdimens}
Q=Q_{0}+\triangle Q&=&\frac{v^2}{M^2}\left(\tilde
Q_{0}+\frac{e^2v^2}{M^2}\triangle\tilde Q\right),\\
\label{Edimens} E=E_{0}+\triangle E&=&\frac{v^2}{M}\left(\tilde
E_{0}+\frac{e^2v^2}{M^2}\triangle\tilde E\right)
\end{eqnarray}
with $\tilde Q_{0}$, $\tilde E_{0}$, $\triangle\tilde Q$ and
$\triangle\tilde E$ being dimensionless functions depending on
$\tilde\omega$ and $\frac{m^2}{M^2}$ only. This suggests that the
parameter
\begin{equation}\label{alpha2}
\alpha_{2}=\frac{e^2v^2}{M^2}
\end{equation}
in this model is such that $\alpha(\omega)\sim\alpha_{2}$. It is
confirmed by the fact that, as can be shown from (\ref{lin1}) and
(\ref{lin2}) using (\ref{potential2}) and (\ref{fdimens}),
solutions for the fields $g$ and $\varphi$ can be expressed in the
form
\begin{eqnarray}\label{gdimens}
g=M\alpha_{2}\tilde g,\\ \label{varphidimens}
\varphi=v\alpha_{2}\tilde\varphi.
\end{eqnarray}
where the dimensionless functions $\tilde g$ and $\tilde\varphi$
depend only on $\tilde\omega$, $\frac{m^2}{M^2}$ and $\tilde r$
and do not depend on $v$ and $e$.

As in the previous case, Eq.~(\ref{alpha2}) clearly shows that the
linear approximation can be used if not only $e^2\ll 1$ holds, but
if $\alpha(\omega)\ll 1$ holds, too. Indeed, even for a very small
value of $e^{2}$, the relation $\frac{v^2}{M^2}$ can be large
enough to make the use of the linear approximation impossible (as
we will see below, for $\alpha_{2}=0.001$ the linear approximation
does not work well enough for all values of $\omega$, whereas such
value of $\alpha_{2}$ can be obtained by choosing $v=M$ and
$e^{2}=0.001$, which looks small enough). On the other hand, for
larger values of $e^{2}$, the value of $\frac{v^2}{M^2}$ can be
chosen to be rather small to make $\alpha(\omega)\ll 1$.

\begin{figure}[!h]
\begin{center}
\includegraphics[width=0.88\linewidth]{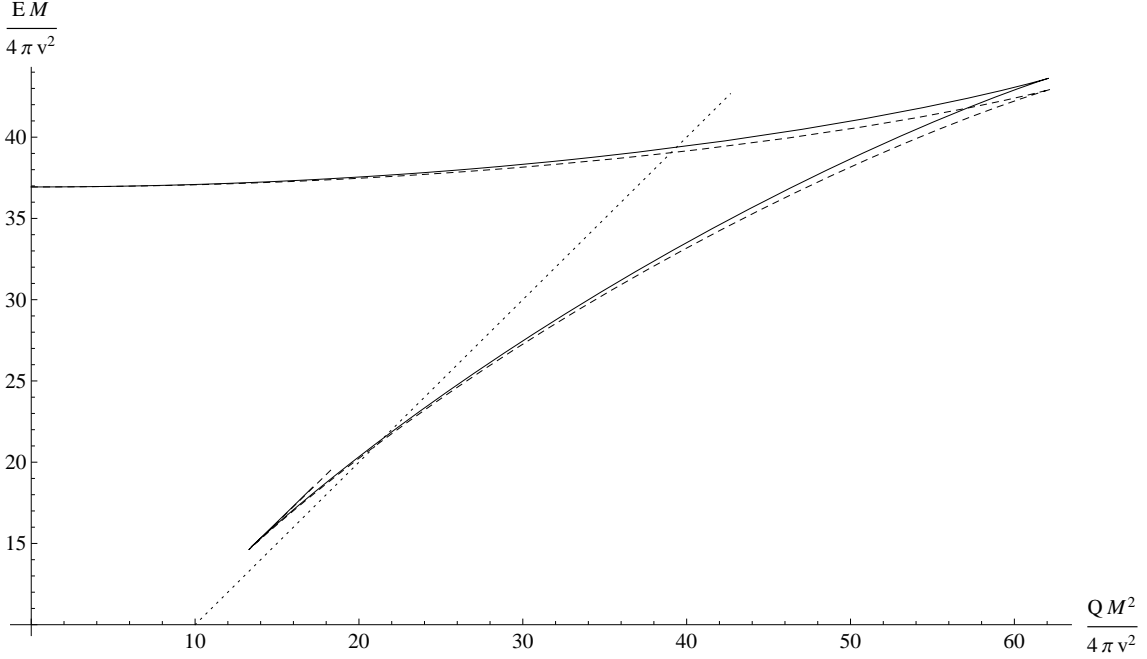}
\caption{$E(Q)$ for the gauged (solid line) and nongauged (dashed
line) cases. The dotted line stands for free scalar particles of
mass $M$ at rest. Here, $m^2<0$, $\frac{|m|}{M}=0.6$,
$\alpha_{2}=0.001$, and $0\le\tilde\omega\le 0.99$.}\label{EQ}
\end{center}
\end{figure}

For a numerical analysis, we choose the case $m^{2}<0$, which is,
in our opinion, the most interesting for illustrative purposes. In
Fig.~\ref{EQ}, one can see an example of the $E(Q)$ diagram for
the gauged Q-ball in our model. This diagram was plotted using
formulas (\ref{mainresultlin0}) and (\ref{mainresultlin}).

Let us discus the properties of the gauged Q-balls at hand. Again
we see from Fig.~\ref{EQ} that the energy of the gauged Q-ball is
larger than the energy of the corresponding nongauged Q-ball for
the same values of charge. One also sees that there are two cusps
on the $E(Q)$ diagram for the gauged case as well as for the
nongauged case. There are locally minimal charges $Q^{min}$ and
locally maximal charges $Q^{max}$, which correspond to these
cusps. For the nongauged case with $\frac{|m|}{M}=0.6$, the charge
is locally maximal at $\tilde\omega\approx 0.2846$, whereas it is
locally minimal at $\tilde\omega\approx 0.9426$. We calculated
numerically the values of $\triangle Q$ for these values of
$\tilde\omega$. According to (\ref{cusppos}), we have
\begin{eqnarray}\nonumber
\frac{M^2}{4\pi v^2}(Q^{max}-Q_{0}^{max})\approx -0.421\,\alpha_{2},\\
\nonumber \frac{M^2}{4\pi v^2}(Q^{min}-Q_{0}^{min})\approx
0.139\,\alpha_{2}.
\end{eqnarray}
Of course, the difference between $Q^{max}$ and $Q_{0}^{max}$ can
not be seen in Fig.~\ref{EQ} by the naked eye because of the small
value of the parameter $\alpha_{2}$. Contrary to the case of the
previous model, here $Q^{max}<Q_{0}^{max}$ and
$Q^{min}>Q_{0}^{min}$. The latter relations are not universal even
in the model under consideration. For example, for
$\frac{|m|}{M}=1.3$,
\begin{eqnarray}\nonumber
\frac{M^2}{4\pi v^2}(Q^{max}-Q_{0}^{max})\approx 0.0009\,\alpha_{2},\\
\nonumber \frac{M^2}{4\pi v^2}(Q^{min}-Q_{0}^{min})\approx
0.0056\,\alpha_{2},
\end{eqnarray}
i.e., now $Q^{max}>Q_{0}^{max}$.

\begin{figure}[!h]
\begin{center}
\includegraphics[width=0.9\linewidth]{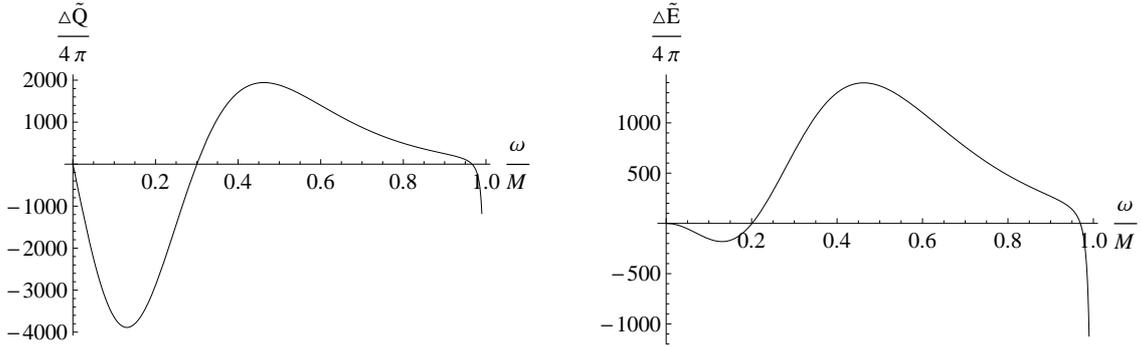}
\caption{$\triangle\tilde Q$ (left plot) and $\triangle\tilde E$
(right plot) for $m^2<0$, $\frac{|m|}{M}=0.6$ and
$0\le\tilde\omega\le 0.99$.}\label{EQ-corr2}
\end{center}
\end{figure}

We also present the plots of $\triangle\tilde Q$ and
$\triangle\tilde E$; see Fig.~\ref{EQ-corr2}.
Again, one sees from these plots that the corrections $\triangle
Q$ and $\triangle E$ can be negative or positive for a given
$\omega$.

\begin{figure}[ht]
\begin{center}
\includegraphics[width=0.9\linewidth]{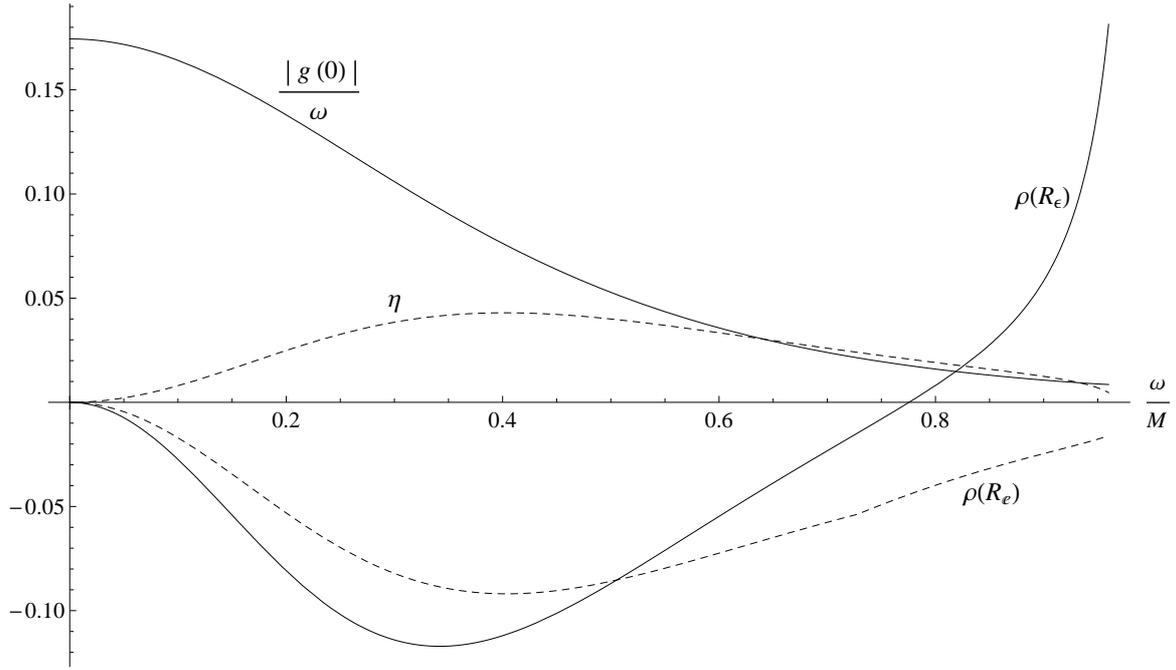}
\caption{$\frac{|g(0)|}{\omega}$, $\eta$ and $\rho$ for $m^2<0$,
$\frac{|m|}{M}=0.6$, $\alpha_{2}=0.001$, and $0\le\tilde\omega\le
0.96$.}\label{gtoomega}
\end{center}
\end{figure}

The plots of the parameters $\frac{|g(0)|}{\omega}$,
$\eta=\frac{|2\triangle E-\omega\triangle Q|}{2\omega Q_{0}}$, and
$\rho(r_{i})=\frac{|g(0)|}{f_{0}(r_{i})}\frac{df_{0}(r_{i})}{d\omega}$,
which are necessary for checking the validity of the linear
approximation, are presented in Fig.~\ref{gtoomega}.
All these parameters were calculated numerically. We calculated
$\rho(r)$ at two points: the first one is defined by
$f_{0}(R_{\textrm{e}},\omega)=\textrm{e}^{-1}f_{0}(0,\omega)$,
whereas the second point, $R_{\epsilon}$, is defined by
(\ref{defRhat}) with $\epsilon=10^{-2}$ and corresponds to the
radius of the inner region. Both $R_{\textrm{e}}$ and
$R_{\epsilon}$ depend on $\omega$; see Appendix~D for details
concerning the calculation of $R_{\textrm{e}}$ and $R_{\epsilon}$.

\begin{figure}[!ht]
\begin{center}
\includegraphics[width=0.9\linewidth]{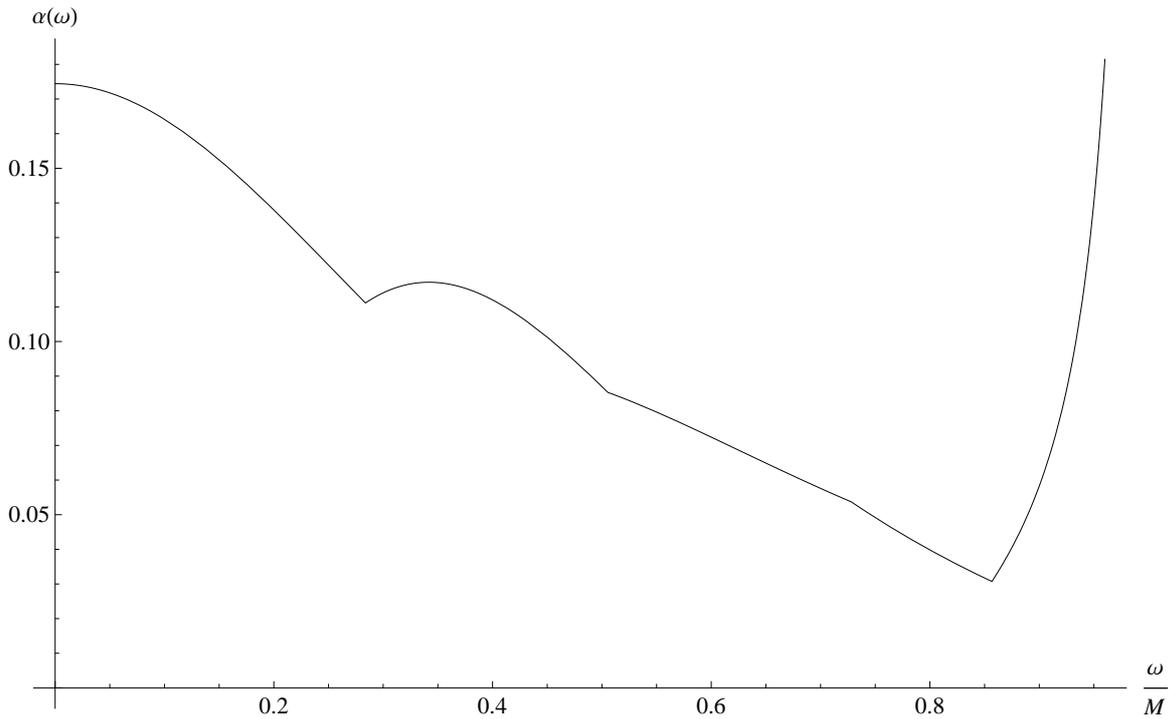}
\caption{$\alpha(\omega)$ for $m^2<0$, $\frac{|m|}{M}=0.6$,
$\alpha_{2}=0.001$, and $0\le\tilde\omega\le
0.96$.}\label{alphafig}
\end{center}
\end{figure}

We see from Fig.~\ref{gtoomega} that all the parameters depend on
$\omega$ in different ways. This explicit example confirms that in
order to check the validity of the linear approximation in the
general case (in which the dependence of these parameters on
$\omega$ is very complicated or can not be obtained analytically),
it is better to estimate all the parameters
$\frac{|g(0)|}{\omega}$, $\eta=\frac{|2\triangle E-\omega\triangle
Q|}{2\omega Q_{0}}$, and
$\rho(r_{i})=\frac{|g(0)|}{f_{0}(r_{i})}\frac{df_{0}(r_{i})}{d\omega}$
or to calculate the function
$\alpha(\omega)=\underset{i}{\textrm{max}}\left\{\frac{|g(0)|}{\omega},\eta,|\rho(r_{i})|\right\}$,
defined by (\ref{expparam}), which is presented in
Fig.~\ref{alphafig} for the set of the model parameters chosen
above.
Fig.~\ref{gtoomega} and Fig.~\ref{alphafig} demonstrate that,
although the use of parameters like $\alpha_{2}$ as
$\omega$-independent small parameters can be convenient for
calculations and for rough estimates, they can not replace the
natural small parameters $\alpha(\omega)$ in the general case.

Of course, the smaller $\alpha_{2}$ is, the wider the region (or
regions) of frequencies $\omega$, in which the linear
approximation works, is. Meanwhile, the chosen set of the
parameters, for which Fig.~\ref{gtoomega} and Fig.~\ref{alphafig}
were plotted, is very useful for illustrative purposes. Based on
these reasons, as well as to make the differences between the
gauged and nongauged cases visible by the naked eye, we keep
Fig.~\ref{EQ} as it is, although, according to
Fig.~\ref{alphafig}, the linear approximation works well enough
only in the vicinity of $\tilde\omega\approx 0.85$ for
$\alpha_{2}=0.001$.

Now, let us turn to the discussion of stability of gauged Q-balls
in this model. We will focus on the lowest branch in
Fig.~\ref{EQ}, for which $\frac{d^2E}{dQ^2}<0$. The results of
Section~3 imply that Q-balls from the lowest branch are stable
against fission. We also see that the part of this lowest branch
lies below the line $E=MQ$ standing for free scalar particles of
mass $M$. This means that, at least in the absence of fermions,
Q-balls from this part of the branch are stable with respect to
decay into free particles.

The last type of stability, which could be discussed here, is the
classical stability. Of course, we may wonder that the main
properties of gauged and nongauged Q-balls are similar at least in
the case in which the parameter $\alpha(\omega)$ is rather small.
So, if we suppose that the classical stability criterion for
ordinary Q-balls \cite{Friedberg:1976me}, which states that
Q-balls for which
$\frac{dQ}{d\omega}=\left(\frac{d^2E}{dQ^2}\right)^{-1}<0$ (for
$Q>0$, $\omega>0$) holds are classically stable, is valid for
gauged Q-balls, then the gauged Q-balls from the lowest branch are
also classically stable. In such a case, the part of the lowest
branch, which lies below the line $E=MQ$, consists of absolutely
stable Q-balls. But we can not justify that the classical
stability criterion for ordinary Q-balls works for gauged Q-balls
too, so we consider the Q-balls from the part of the lowest
branch, which lies below the $E=MQ$ line, as stable against
fission and against decay into free scalar particles only.
According to Fig.~\ref{EQ}, for the given values of the model
parameters ($m^2<0$, $\frac{|m|}{M}=0.6$, $\alpha_{2}=0.001$),
such Q-balls have charges in the range
$21.5\lesssim\frac{M^2}{4\pi v^2}Q\lesssim 61.8$ and energies in
the range $21.5\lesssim\frac{M}{4\pi v^2}E\lesssim 43.5$.

As for the previous model, for completeness below we present an
explicit solution for the fields $g$ and $\varphi$ in this model.
The solution for the field $g$ can be obtained directly from
(\ref{Rosensol}) with (\ref{backsol1}) and (\ref{backsol2}),
whereas solution for the field $\varphi$ was obtained by means of
the method of variation of parameters. We present it for the
simplest case $m=0$. This solution, satisfying
$\frac{dg}{dr}\bigl|_{r=0}=0$, $g|_{r\to\infty}=0$,
$\frac{d\varphi}{dr}\bigl|_{r=0}=0$, $\varphi|_{r\to\infty}=0$,
and the nonstandard matching conditions at $r=R$ (because
$\frac{d^2V}{df^2}\bigl|_{f=v}\sim\delta(r-R)$), has the form for
the gauge field
\begin{eqnarray}\label{linsolbegin}
g(r<R)=g_{<}(r)&=&C_{1}\left(\ln(\omega r)-\textrm{Ci}(2\omega
r)+\frac{\sin(2\omega r)}{2\omega r}\right)+C_{2},\\
g(r>R)=g_{>}(r)&=&\frac{C_{3}}{r}+C_{4}\left(\frac{\textrm{e}^{-2\sqrt{M^2-\omega^2}\,r}}{2\sqrt{M^2-\omega^2}\,r}-\textrm{E}_{1}(2\sqrt{M^2-\omega^2}\,r)\right),
\end{eqnarray}
where
\begin{eqnarray}
\textrm{Ci}(y)&=&-\int\limits_{y}^{\infty}\frac{\cos(t)}{t}dt,
\end{eqnarray}
\begin{eqnarray}
C_{1}=C_{1}(\omega)&=&e^2v^2\omega R^2\frac{1}{\sin^{2}(\omega
R)},\\ \nonumber C_{2}=C_{2}(\omega)&=&-e^2v^2\omega
R^2\left.\biggl(2\,\textrm{e}^{2\sqrt{M^2-\omega^2}R}\textrm{E}_{1}(2\sqrt{M^2-\omega^2}R)\right.\\
&+&\left.\frac{-\textrm{Ci}(2\omega R)+\ln(\omega
R)+1}{\sin^{2}(\omega R)}\right),\\
C_{3}=C_{3}(\omega)&=&-e^2v^2\omega R^2\left(\frac{M^2}{\omega^{2}\sqrt{M^2-\omega^2}}+\frac{R}{\sin^{2}(\omega R)}\right),\\
C_{4}=C_{4}(\omega)&=&e^2v^2\omega
R^2\left(2\,\textrm{e}^{2\sqrt{M^2-\omega^2}R}\right),
\end{eqnarray}
and for the scalar field
\begin{eqnarray}\nonumber
\varphi(r<R)&=&B\frac{\sin(\omega r)}{r}+\frac{\sin(\omega
r)}{\omega r}\int\limits_{0}^{r}G_{<}(t)\cos(\omega t)dt\\  &-&
\frac{\cos(\omega r)}{\omega
r}\int\limits_{0}^{r}G_{<}(t)\sin(\omega t)dt,\\
\nonumber
\varphi(r>R)&=&A\frac{\textrm{e}^{-\sqrt{M^2-\omega^2}\,r}}{r}-\frac{\textrm{e}^{\sqrt{M^2-\omega^2}\,r}}{2\sqrt{M^2-\omega^2}\,r}
\int\limits_{r}^{\infty}G_{>}(t)\textrm{e}^{-\sqrt{M^2-\omega^2}\,t}dt\\
&-&\frac{\textrm{e}^{-\sqrt{M^2-\omega^2}\,r}}{2\sqrt{M^2-\omega^2}\,r}
\int\limits_{R}^{r}G_{>}(t)\textrm{e}^{\sqrt{M^2-\omega^2}\,t}dt,
\end{eqnarray}
where
\begin{eqnarray}
G_{<}(r)&=&-2\omega r g_{<}(r)f_{0}^{<}(r),\\
G_{>}(r)&=&-2\omega r g_{>}(r)f_{0}^{>}(r),
\end{eqnarray}
\begin{eqnarray}
B=B(\omega)&=&\frac{1}{D}F_{1}\frac{\textrm{e}^{\sqrt{M^2-\omega^2}R}}{\sin(\omega R)}-\frac{F_{2}}{\omega}+\frac{F_{3}}{\omega^2 R},\\
A=A(\omega)&=&\frac{\textrm{e}^{\sqrt{M^2-\omega^2}R}}{D}\left(F_{1}\textrm{e}^{\sqrt{M^2-\omega^2}R}\left(1+\frac{D}{2\sqrt{M^2-\omega^2}}\right)+F_{3}\frac{M^2\sin(\omega
R)}{\omega^2}\right),
\end{eqnarray}
\begin{eqnarray}
D=D(\omega)&=&\frac{M^2R}{1+R\sqrt{M^2-\omega^2}},\\
F_{1}=F_{1}(\omega)&=&\int\limits_{R}^{\infty}G_{>}(t)\textrm{e}^{-\sqrt{M^2-\omega^2}t}dt,\\
F_{2}=F_{2}(\omega)&=&\int\limits_{0}^{R}G_{<}(t)\cos(\omega t)dt,\\
\label{linsolend}
F_{3}=F_{3}(\omega)&=&\int\limits_{0}^{R}G_{<}(t)\sin(\omega t)dt.
\end{eqnarray}
We see that, even for the very simple background solution
(\ref{backsol1}), (\ref{backsol2}), the solution for $g$ and
$\varphi$ appears to be complicated. Contrary to the case of
Model~1, the solution (\ref{linsolbegin})--(\ref{linsolend}) can
not be factorized. It should be mentioned that the double
integrals in (\ref{linsolbegin})--(\ref{linsolend}) (the functions
$\textrm{Ci}(y)$ and $\textrm{E}_{1}(y)$ have integral
representations themselves) in principle can be transformed to the
form containing only integrals of one variable by performing
integration by parts (corresponding calculations are
straightforward, though rather tedious); this is possible only
because of the simplicity of the background solution
(\ref{backsol1}), (\ref{backsol2}).

\begin{figure}[ht]
\begin{center}
\includegraphics[width=0.97\linewidth]{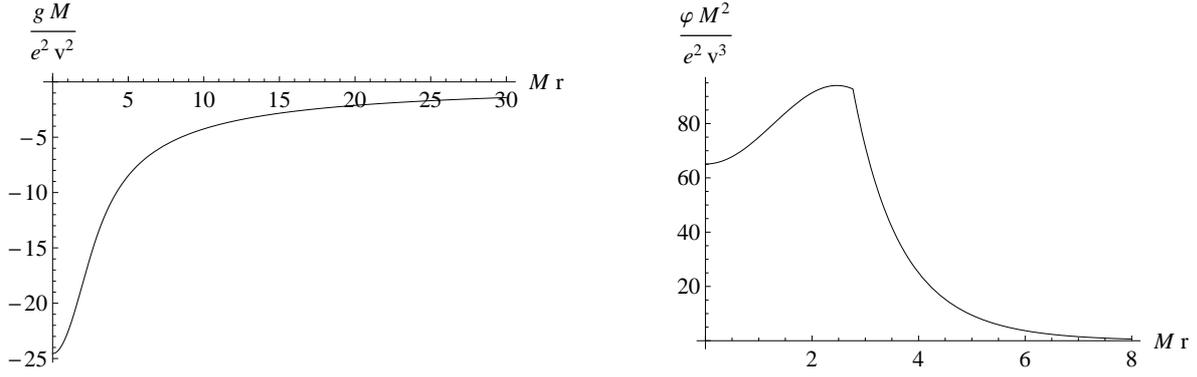}
\caption{Solutions for the fields $g(r)$ (left plot) and
$\varphi(r)$ (right plot). Here, $m=0$ and
$\tilde\omega=0.8$.}\label{sol1}
\end{center}
\end{figure}
\begin{figure}[ht]
\begin{center}
\includegraphics[width=0.97\linewidth]{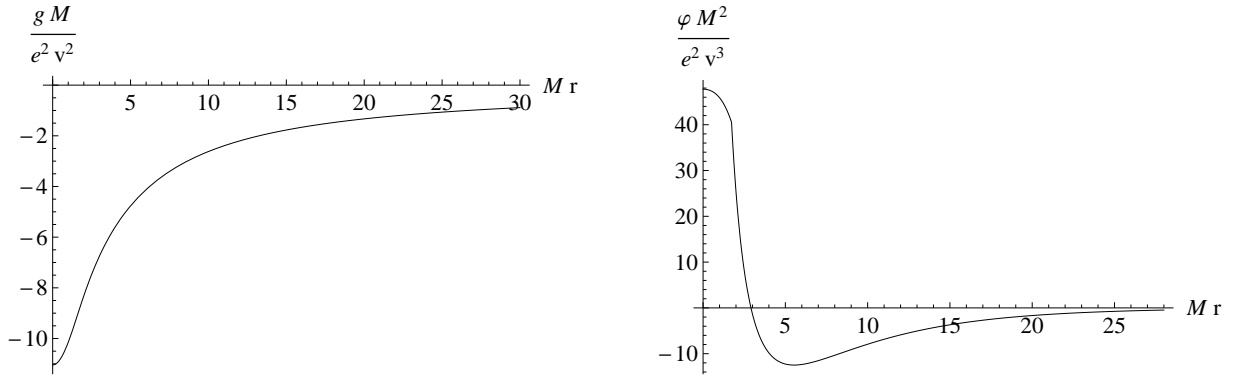}
\caption{Solutions for the fields $g(r)$ (left plot) and
$\varphi(r)$ (right plot). Here, $m=0$ and
$\tilde\omega=0.99$.}\label{sol2}
\end{center}
\end{figure}

The plots of this solution for the fields $g$ and $\varphi$ are
presented in Figs.~\ref{sol1}~and~\ref{sol2}. One can see the
breaks on the curves corresponding to the field $\varphi$. This is
an artefact of linearization in the theory with potential
(\ref{potential2}), which also contains a break (recall that
$\frac{d^2V}{df^2}\bigl|_{f=v}\sim\delta(r-R)$). Of course, the
break in potential (\ref{potential2}) can be regularized, leading
to the smooth behavior of $\varphi$ at $r=R$.

One also sees that for different values of $\tilde\omega$
solutions for the field $\varphi$ have different form: in the
first case, $\varphi$ increases at small $r$, whereas in the
second case, it decreases at small $r$ starting from $r=0$. It
should be noted that, contrary to what we have in Model~1 (see
(\ref{Fvarphi})), in both cases, the solution for the field
$\varphi$ crosses the axis $Mr$ and then tends to zero from below.
It is clearly seen in Fig.~\ref{sol2}; as for the case presented
in Fig.~\ref{sol1}, the solution crosses the axis $Mr$ at
$Mr\approx 17.38$, which is out of range of the presented plot.
Solutions for other values of the model parameters have the form
similar either to the solution presented in Fig.~\ref{sol1} or to
the solution presented in Fig.~\ref{sol2}.

Finally, we would like to note that, as it can be checked
explicitly, the relation $\frac{|\varphi(r)|}{f_{0}(r)}$ grows
logarithmically with $r$ in this model (see also Appendix~C). But,
again, this growth is very slow relative to the exponential fall
of $f_{0}$ and $\varphi$, and, analogously to the previous case,
this confirms that with an appropriate choice of the model
parameters (including $\omega$) and with $\alpha(\omega)\ll 1$ the
use of the linear approximation is fully justified.

\section{Conclusion}
In the present paper, we studied some general properties of $U(1)$
gauged Q-balls. In particular, we showed that, as in the case of
ordinary nongauged Q-balls, the relation $\frac{dE}{dQ}=\omega$
also holds for $U(1)$ gauged Q-balls. Based on this result and
using the fact that $\omega<M$ in theories admitting the existence
of free scalar particles of mass $M$, we demonstrated that the
statement about the existence of the maximal charge of stable
gauged Q-balls, presented in \cite{Lee:1988ag}, was obtained by
means of the erroneous inequality $\frac{dE}{dQ}\ge M$ and thus
can not be considered as correct.

We also presented a powerful method for analyzing gauged Q-balls
in the case in which the back-reaction of the gauge field on the
scalar field is small. Provided a nongauged (background) Q-ball
solution $f_{0}(r,\omega)$, for a given value of the coupling
constant $e$, the strength of the back-reaction of the gauge field
can be estimated by calculating the parameter $\alpha(\omega)$
defined by (\ref{expparam}), which depends on $\omega$ and the
background solution $f_{0}(r,\omega)$ only. This parameter is
proportional to $e^2$ in general but does not coincide with it. We
have shown that our results can be used not if only the inequality
$e^{2}\ll 1$ holds but if the overall parameter $\alpha(\omega)$
is also rather small to ensure the validity of the linear
approximation (in principle, the smallness of $\alpha(\omega)$
does not exclude the cases in which $e$ is not small). The main
parameters of gauged Q-balls in such a theory
--- the charge and the energy --- can also be calculated using the
background solution $f_{0}(r,\omega)$ only (using Eqs.
(\ref{mainresultlin0}), (\ref{mainresultlin}) or the even simpler
Eq. (\ref{EgE0-main})), whereas an explicit solution to the system
of linearized equations of motion is not necessary at all.

The obtained results were illustrated by the examples of two
exactly solvable models proving the efficiency of the proposed
method --- indeed, even for the very simple background solution
(\ref{backsol1}), (\ref{backsol2}), the explicit analytic solution
for gauged Q-ball (\ref{linsolbegin})--(\ref{linsolend}), which
was obtained in the linear approximation in $\varphi$ and $g$,
appears to be rather complicated and its derivation (at least for
the field $\varphi$) is more bulky than the analytical evaluation
of integral (\ref{Ifinal}) for (\ref{backsol1}) and
(\ref{backsol2}). Without
(\ref{mainresultlin0})--(\ref{mainresultlin1}) or
(\ref{EgE0-main}), evaluation of (\ref{charge}) and (\ref{energy})
does not seem to be a simple task, taking into account the
necessity to solve numerically (in the general case) the
differential equation (\ref{lin2}) to get a solution for the field
$\varphi$. Obviously, it is a much more complicated task than the
evaluation of the double integral in (\ref{Ifinal}), even with a
background nongauged solution $f_{0}(r,\omega)$ obtained
numerically.

We hope that the results presented in this paper can be useful for
the future research in this area.

\section*{Acknowledgements}
The authors are grateful to M.V.~Libanov, V.A.~Rubakov,
S.V.~Troitsky, and I.P.~Volobuev for valuable discussions. The
work was supported by RFBR grant 14-02-31384. The work of E.Y.N.
was supported in part by grant NS-2835.2014.2 of the President of
Russian Federation and by RFBR grant 13-02-01127a. The work of
M.N.S. was supported in part by grant NS-3042.2014.2 of the
President of Russian Federation and by RFBR grant
12-02-93108-CNRSL-a.

\section*{Appendix A}
Let
$\frac{E(Q_{min})}{Q_{min}}=\hat\omega>\omega_{min}=\frac{dE}{dQ}\bigl|_{Q=Q_{min}}$.
In this case, the function $E_{aux}(Q)$ can be chosen as
\begin{eqnarray}\nonumber
E_{aux}(Q)&=&(2\hat\omega-\omega_{min})Q+\frac{Q^{2}}{Q_{min}}(\omega_{min}-\hat\omega),\qquad
Q<Q_{min},\\ \nonumber E_{aux}(Q)&=&E(Q),\hspace{5.75cm} Q\ge
Q_{min}.
\end{eqnarray}
One can check that
\begin{eqnarray}\nonumber
&&E_{aux}(0)=0,\\ \nonumber &&E_{aux}(Q_{min})=E(Q_{min}),
\end{eqnarray}
\begin{eqnarray}\nonumber
&&\frac{dE_{aux}(Q)}{dQ}\biggl|_{Q=Q_{min}}=\frac{dE(Q)}{dQ}\biggl|_{Q=Q_{min}},\\
\nonumber &&\frac{dE_{aux}(Q)}{dQ}=2(\hat
\omega-\omega_{min})\left(1-\frac{Q}{Q_{min}}\right)+\omega_{min}>0\qquad
\textrm{for}\quad Q\le Q_{min},\\ \nonumber
&&\frac{d^{2}E_{aux}(Q)}{dQ^{2}}=\frac{2}{Q_{min}}(\omega_{min}-\hat
\omega)<0,
\end{eqnarray}
i.e., all the necessary conditions are fulfilled. Since
$E_{aux}(0)=0$, inequality (\ref{q-decay2}) holds for $Q_{1},
Q_{2}\ge 0$ and, consequently, for $Q_{1}, Q_{2}\ge Q_{min}$. For
more details, see \cite{Gulamov:2013ema}.

\section*{Appendix B}
Let us consider Eqs.~(\ref{lin1}) and (\ref{lin2}) and represent
the coupling constant $e$ in (\ref{lin1}) as $e=\gamma e'$, where
$\gamma>0$ is a constant. Let us define $f'_{0}=\gamma f_{0}$,
$\varphi'=\gamma\varphi$. With these notations, Eqs.~(\ref{lin1})
and (\ref{lin2}) can be rewritten as
\begin{eqnarray}\label{lin1append}
\Delta g-2\,{e'}^2\omega {f'_{0}}^2=0,\\ \label{lin2append}
\Delta\varphi'+\omega^2\varphi'+2\omega
gf'_{0}-\frac{1}{2}\frac{d^2V'(f')}{d{f'}^2}\biggl|_{f'=f'_{0}}\varphi'=0,
\end{eqnarray}
where $V'(f')=\gamma^{2}V\left(\frac{f'}{\gamma}\right)$.
Eqs.~(\ref{lin1append}) and (\ref{lin2append}) have the same form
as Eqs.~(\ref{lin1}) and (\ref{lin2}), but now with the coupling
constant $e'$ instead of $e$ and with the scalar field potential
that differs from the one in (\ref{lin2}). Meanwhile, in fact, the
system of equations remains the same --- we have only changed the
variables. This simple argumentation shows that not only the
coupling constant $e$ defines whether the linear approximation can
be used, but it is the coupling constant {\em together} with the
parameters of the scalar field potential.

\section*{Appendix C}
Let us show that $\frac{|\varphi(r)|}{f_{0}(r)}$ grows
logarithmically with $r$ for potentials satisfying
$$
\frac{dV}{df}\biggl|_{f=0}=0,\qquad
\frac{1}{2}\frac{d^{2}V}{df^{2}}\biggl|_{f=0}=M^{2}.
$$
It is clear that for $\sqrt{M^{2}-\omega^{2}}\,r\gg 1$ the
background solution in such a model has the form
$f_{0}(r)\sim\frac{\textrm{e}^{-\sqrt{M^{2}-\omega^{2}}\,r}}{r}$,
whereas $g(r)\sim\frac{1}{r}$. So, equation (\ref{lin2}) can be
written as
\begin{eqnarray}\label{lin2infty}
\Delta\varphi+(\omega^2-M^2)\varphi=C\,\frac{\textrm{e}^{-\sqrt{M^{2}-\omega^{2}}\,r}}{r^{2}},
\end{eqnarray}
for $\sqrt{M^{2}-\omega^{2}}\,r\gg 1$, where $C$ is a constant (in
fact, $C=C(\omega)$, but it is not important for the present
calculation). Now, we define $\psi=r\varphi$ and rewrite
(\ref{lin2infty}) as
\begin{eqnarray}\label{lin2infty1}
\frac{d^{2}\psi}{dr^{2}}+(\omega^2-M^2)\psi=C\,\frac{\textrm{e}^{-\sqrt{M^{2}-\omega^{2}}\,r}}{r}.
\end{eqnarray}
Eq.~(\ref{lin2infty1}) can easily be solved by means, say, of the
method of variation of parameters. Its solution (such that
$\psi(r)|_{r\to\infty}\to 0$) takes the form
\begin{equation}
\psi=-\frac{C}{2\sqrt{M^{2}-\omega^{2}}}\left(\textrm{e}^{-\sqrt{M^{2}-\omega^{2}}\,r}\ln(\xi\,
r)+\textrm{e}^{\sqrt{M^{2}-\omega^{2}}\,r}\int\limits_{r}^{\infty}\frac{\textrm{e}^{-2\sqrt{M^{2}-\omega^{2}}\,z}}{z}dz\right),
\end{equation}
where $\xi$ is a constant. Recalling that
$\varphi=\frac{\psi}{r}$, in the leading order we get
$$\frac{|\varphi(r)|}{f_{0}(r)}\biggl|_{r\to\infty}\sim \ln(\xi\,
r).$$

\section*{Appendix D}
The radius $R_{\textrm{e}}=R_{\textrm{e}}(\omega)$ is defined by
the relation $f_{0}(R_{\textrm{e}})=\textrm{e}^{-1}f_{0}(0)$.
Using Eqs.~(\ref{backsol1}),~(\ref{backsol2}), and (\ref{Rdef}),
it is not difficult to show that $R_{\textrm{e}}$ satisfies the
following equations:
\begin{eqnarray}\label{firstRe}
\frac{\sqrt{\omega^{2}-m^{2}}R_{\textrm{e}}}{\sin(\sqrt{\omega^{2}-m^{2}}R_{\textrm{e}})}=\textrm{e},\qquad
\textrm{for}\qquad R_{\textrm{e}}<R,\\ \label{secondRe}
\frac{\sqrt{M^{2}-m^{2}}R_{\textrm{e}}\textrm{e}^{\sqrt{M^{2}-\omega^{2}}R_{\textrm{e}}}}{\textrm{e}^{\sqrt{M^{2}-\omega^{2}}R}}=\textrm{e},\qquad
\textrm{for}\qquad R_{\textrm{e}}\ge R.
\end{eqnarray}
At first, we solved numerically Eq.~(\ref{firstRe}) for a given
$\omega$. The result satisfying $R_{\textrm{e}}<R=R(\omega)$ was
accepted; the result satisfying $R_{\textrm{e}}>R=R(\omega)$ was
rejected, and the solution to Eq.~(\ref{secondRe}) was accepted as
the value of $R_{\textrm{e}}$.

The coordinate $R_{\epsilon}$ is defined by Eq.~(\ref{defRhat}).
Suppose that $R_{\epsilon}=R_{\epsilon}(\omega)>R(\omega)$ for
$m^2<0$, $\frac{|m|}{M}=0.6$, and $\epsilon=10^{-2}$. In this
case, Eq.~(\ref{defRhat}) gives
\begin{equation}
R_{\epsilon}=R+\frac{1}{2\sqrt{M^{2}-\omega^{2}}}\ln\left(\frac{(\epsilon+1)(\omega^{2}-m^{2})}{\epsilon\,
(M^{2}-m^{2})(1+R\sqrt{M^{2}-\omega^{2}})}\right).
\end{equation}
It can be checked that the equality
$R_{\epsilon}(\omega)>R(\omega)$ indeed holds for the chosen set
of the parameters.


\begin{thebibliography}{99}
\bibitem{Rosen0}
G. ~Rosen, J.\ Math.\ Phys. {\bf 9} (1968) 996.

\bibitem{Coleman:1985ki}
  S.~R.~Coleman,
   Nucl.\ Phys.\ B {\bf 262} (1985) 263
   [Erratum-ibid.\ B {\bf 269} (1986) 744].

\bibitem{Lee:1988ag}
  K.~-M.~Lee, J.~A.~Stein-Schabes, R.~Watkins and L.~M.~Widrow,
    Phys.\ Rev.\ D {\bf 39} (1989) 1665.

\bibitem{Rosen}
G.~Rosen, J. Math. Phys. {\bf 9} (1968) 999.

\bibitem{Lee:1991bn}
  C.~H.~Lee and S.~U.~Yoon,
    Mod.\ Phys.\ Lett.\ A {\bf 6} (1991) 1479.

\bibitem{Friedberg:1976me}
  R.~Friedberg, T.~D.~Lee and A.~Sirlin,
    Phys.\ Rev.\ D {\bf 13} (1976) 2739.

\bibitem{Arodz:2008nm}
  H.~Arodz and J.~Lis,
  Phys.\ Rev.\ D {\bf 79} (2009) 045002.

\bibitem{BF}
V.~Benci and D.~Fortunato, J. Math. Phys. {\bf 52} (2011) 093701.

\bibitem{BF1}
V.~Benci and D.~Fortunato, arXiv:1212.3236 [math-ph].

\bibitem{Dzhunushaliev:2012zb}
  V.~Dzhunushaliev and K.~G.~Zloshchastiev,
   Central Eur.\ J.\ Phys.\  {\bf 11} (2013) 325.

\bibitem{Rosen:1969ay}
  G.~Rosen,
   Phys.\ Rev.\  {\bf 183} (1969) 1186.

\bibitem{Loginov:2012zz}
  A.~Y.~Loginov,
    J.\ Exp.\ Theor.\ Phys.\  {\bf 114} (2012) 48.

\bibitem{Gulamov:2013ema}
  I.~E.~Gulamov, E.~Y.~Nugaev and M.~N.~Smolyakov,
    Phys.\ Rev.\ D {\bf 87} (2013) 085043.

\bibitem{MarcVent}
G.~C.~Marques and I.~Ventura,
    Phys.\ Rev.\ D {\bf 14} (1976) 1056.

\bibitem{Theodorakis:2000bz}
  S.~Theodorakis,
    Phys.\ Rev.\ D {\bf 61} (2000) 047701.

\bibitem{Derrick}
G.~H.~Derrick, J. Math. Phys. {\bf 5} (1964) 1252.


\end{thebibliography}
\end{document}